\documentclass[a4paper,10pt]{article}
\usepackage{geometry}

\usepackage[T1]{fontenc}
\usepackage[utf8]{inputenc}

\usepackage{amssymb, amsthm, amsmath}
\usepackage{amsmath,amsfonts,amsthm,bm}
\usepackage{authblk}

\usepackage{color}

\usepackage{comment}

\theoremstyle{definition}
\newtheorem{definition}{Definition}

\theoremstyle{plain}
\newtheorem{theorem}{Theorem}
\newtheorem{proposition}[definition]{Proposition}

\newtheorem{lemma}[definition]{Lemma}
\newtheorem{remark}[definition]{Remark}
\newtheorem{corollary}[definition]{Corollary}

\newcommand{\w}{{\rm w}}
\newcommand{\dd}{{\rm d}}
\newcommand{\C}{\mathcal{C}}

\title{On the structure of polynomial ideal codes and\\additive Reed--Solomon codes}
\author{Umberto Mart{\'i}nez-Pe\~{n}as\thanks{umberto.martinez@uva.es} }
\author{Rub{\'e}n Rodr{\'i}guez-Ballesteros\thanks{ruben.rodriguez22@estudiantes.uva.es; rubenrb@iesgalileo.es}}
\affil{IMUVa-Mathematics Research Institute,\\University of Valladolid, Spain}

\date{}

\begin{document}

\maketitle

\begin{abstract}
We study structural properties of polynomial ideal (PI) codes and additive Reed--Solomon (aRS) codes in the folded Hamming metric (equivalently, the Hamming metric over a larger alphabet). We first determine the maximum possible length of PI codes, prove that the dual of every PI code is a generalized PI code, and deduce that PI codes are dually QMDS together with an explicit formula for their weight distribution. We then show that PI codes defined by irreducible polynomials correspond to aRS codes with pairwise nonconjugate evaluation points. For these aRS codes, the previous results on PI codes apply (duality, QMDS property and weight distribution). We show that, asymptotically, aRS codes with nonconjugate evaluation points achieve the same length as PI codes. We also compute the permutation group of such aRS codes. We also extend the family of PI codes to consider arbitrary polynomial modules (not necessarily pairwise coprime), and study when these and general aRS codes satisfy the QMDS and dually QMDS properties.


\end{abstract}

\section{Introduction}
Codes in the folded Hamming metric may be viewed equivalently as codes endowed with the ordinary Hamming metric over an extension alphabet, while retaining linearity only over a prescribed subfield. This point of view includes additive codes and provides a natural framework in which the dimension over the base field need not be a multiple of the alphabet extension degree. In \cite{martinez2025linear} we developed the basic theory of such codes and introduced quasi-MDS (QMDS) codes, which lie in between MDS and almost MDS codes \cite{faldum} (their dimensions are larger than those of almost MDS codes for a given minimum distance, see \cite[Remark 10]{martinez2025linear}). Since we showed in \cite{martinez2025linear} that the dual of a QMDS code need not be QMDS, we also introduced the concept of dually QMDS codes (i.e., codes that themselves and their duals are both QMDS). 

Independently, QMDS codes were introduced in \cite{ball2024griesmer} under the name fractional MDS codes, and in \cite{yadav} under the name additive MDS codes.

Constructions of QMDS codes and dually QMDS codes can be found in \cite{ball2024griesmer, bartoli, martinez2025linear, pavese, yadav}. Upper bounds on their lengths (relative to their base field sizes) have been given in \cite{ball2024griesmer, martinez2026upper}.

Polynomial ideal (PI) codes were introduced in \cite{bhandari2023ideal} in connection with capacity-achieving list decoding. They generalize several well known families of codes such as classical Reed--Solomon codes \cite{reed-solomon}, folded Reed--Solomon codes \cite{guruswami2008explicit}, multiplicity codes \cite{rosenbloom} and, as we will show here, additive Reed--Solomon codes \cite{yadav}. 

Their definition is particularly well adapted to the folded Hamming metric: their codewords are formed by the residues of polynomials of restricted degree modulo several pairwise coprime polynomials of a common degree. In fact, they are always QMDS \cite[Th. 2]{martinez2025linear}. Despite this simple description, several of their structural properties remain unknown, including their largest possible length, computing their duals in general, proving whether they are always dually QMDS and computing their weight distribution. In this work, we fully describe all of these properties of PI codes. We also introduce the class of polynomial residue (PR) codes, which extend PI codes to arbitrary polynomial modules (not necessarily pairwise coprime) and study when they are QMDS and/or dually QMDS.

Additive Reed--Solomon (aRS) codes, introduced in \cite{yadav}, arise from evaluating polynomials over $\mathbb F_q$ at points of an extension field $ \mathbb{F}_{q^r} $. They are therefore $\mathbb F_q$-linear but need not be linear over their alphabet, $ \mathbb{F}_{q^r} $. We will show that there is a large intersection between both families of codes: when the evaluation points belong to pairwise distinct Frobenius conjugacy classes of degree $ r $, aRS codes are linearly isometric (i.e., equivalent) to PI codes with irreducible defining polynomials. In particular, all of the properties that we study for PI codes directly apply to this subfamily of aRS codes. Regarding aRS codes with pairwise nonconjugate evaluation points, we will also give an explicit formula for their duals, show that they asymptotically achieve the maximum possible length of PI codes and compute their permutation group. Finally, we study when general aRS codes are QMDS and dually QMDS in the cases when they are not equivalent to PI codes.

The organization of the manuscript is as follows. In Section \ref{sec preliminaries}, we provide some preliminaries on the folded Hamming metric and QMDS codes; in Section \ref{sec PI codes}, we study PI codes (their maximum length in Subsection \ref{subsec max length}, their duals and weight distribution in \ref{sec:PI-duality}, and when they are QMDS and/or dually QMDS when the polynomial modules are not necessarily coprime in Subsection \ref{subsec PR codes}); and in Section \ref{sec aRS}, we study aRS codes (the equivalence between the classical Hamming and folded Hamming metrics in Subsection \ref{subsec Hamming and folded isometries}, the intersection between PI and aRS codes in Subsection \ref{subsec intersection}, their permutation group in Subsection \ref{subsec permutation} and aRS codes that are not PI codes in Subsection \ref{subsec aRS not covered}).


\section{Preliminaries} \label{sec preliminaries}
Throughout the manuscript, for integers $ n \geq 1 $ and $a\leq b$ we write
$[a,b]=\{a,a+1,\ldots,b\}$ and $[n]=[1,n]$. We denote by $S_n$ the
symmetric group on $[n]$. We regard $\mathbb{F}_q^{nr}$ as $n$ blocks
of length $r$ over $ \mathbb{F}_q $. The folded Hamming metric is precisely the Hamming metric
on these $n$ blocks, so each block may be viewed either as a symbol of
the alphabet $\mathbb{F}_q^r$ or, after choosing an $\mathbb{F}_q$-basis,
as a symbol of $\mathbb{F}_{q^r}$ (see also Subsection \ref{subsec Hamming and folded isometries}). We recall only the definitions needed
below and refer to \cite{martinez2025linear} for further background.
\begin{definition}\label{folded distance}
For $\mathbf{c} = (\mathbf{c}_1,\ldots,\mathbf{c}_n) \in \mathbb{F}_{q}^{nr} $,  where $ \mathbf{c}_i \in \mathbb{F}_q^r $ for $ i \in [n] $, its folded Hamming weight is defined as $\w_F(\mathbf{c})=\lvert\{i\in[n]:\mathbf{c}_i\neq \mathbf{0} \}\rvert$. We define the folded Hamming distance between $ \mathbf{c},\mathbf{d} \in \mathbb{F}_q^{nr} $ as $ \dd_F(\mathbf{c},\mathbf{d}) = \w_F(\mathbf{c}- \mathbf{d}) $. In general, a code is a subset $ \mathcal{C} \subseteq \mathbb{F}_q^{nr} $. We define the minimum folded Hamming distance of $ \mathcal{C} $ as $ \dd(\mathcal{C}) = \min \{ \dd_F(\mathbf{c},\mathbf{d}) : \mathbf{c},\mathbf{d} \in \mathcal{C}, \mathbf{c} \neq \mathbf{d} \} $. For the zero code, we adopt the convention $\dd(\{ \mathbf{0} \})=n+1$.
\end{definition}

If $ \mathcal{C} \subseteq \mathbb{F}_q^{rn} $ is a nonzero $\mathbb{F}_q$-linear code, then
\[
\dd(\mathcal{C}) = \min \{ \w_F(\mathbf{c}) : \mathbf{c} \in \mathcal{C} \setminus \{ \mathbf{0} \} \}.
\]

\begin{definition}
We say that $\mathcal{C}$ is a code of type $[n,r,k,d]$ when $\mathcal{C}\subseteq\mathbb{F}_q^{nr}$ is $\mathbb{F}_q$-linear, $k=\dim_{\mathbb F_q}(\mathcal C)$ and $d=\dd(\mathcal C)$.
\end{definition}

Later, we will consider weight distributions.

\begin{definition}
The (folded) weight distribution of a linear code $ \mathcal{C} \subseteq \mathbb{F}_q^{nr} $ is defined as the numbers $W_j\left(\mathcal{C}\right)=\lvert\left\{\mathbf{c}\in\mathcal{C}: \text{w}_F(\mathbf{c})=j\right\}\rvert$, for $ j \in [0,n] $.
\end{definition}

We will also need the concept of isometry and equivalence.

\begin{definition}
Two $\mathbb F_q$-linear codes $\mathcal C,\mathcal D\subseteq\mathbb F_q^{nr}$ are equivalent if there exists an $\mathbb F_q$-linear folded-Hamming isometry $\phi:\mathbb F_q^{nr}\to\mathbb F_q^{nr}$ (i.e., $ \w_F(\phi(\mathbf{c})) = \w_F(\mathbf{c}) $, for all $ \mathbf{c} \in \mathbb{F}_q^{nr} $) such that $\phi(\mathcal C)=\mathcal D$.
\end{definition}

The linear isometries of the folded Hamming metric are exactly the
coordinate permutations followed by invertible $\mathbb F_q$-linear
transformations inside each block. The following result is a particular case of \cite[Th. 2]{sr-hamming}. See also \cite[Prop. 5]{martinez2025linear}.

\begin{proposition}[\textbf{\cite{sr-hamming}}]
\label{prop isometries}
Let $\phi:\mathbb F_q^{nr}\longrightarrow\mathbb F_q^{nr}$ be an $\mathbb F_q$-linear vector space isomorphism. Then
$\w_F(\phi(\mathbf c))=\w_F(\mathbf c)$ for all
$\mathbf c\in\mathbb F_q^{nr}$ if and only if there exist invertible
matrices $A_1,\ldots,A_n\in\mathrm{GL}_r(\mathbb F_q)$ and a permutation $\sigma\in S_n$ such that
\[
\phi(\mathbf c_1,\ldots,\mathbf c_n)
=
\big(
\mathbf c_{\sigma(1)}A_1,\ldots,
\mathbf c_{\sigma(n)}A_n
\big),
\]
for all $\mathbf c_1,\ldots,\mathbf c_n\in\mathbb F_q^r$.
\end{proposition}

Although the folded Hamming metric may be viewed as the
ordinary Hamming metric over the alphabet $\mathbb F_{q^r}$,
throughout the paper duality is taken with respect to the
$\mathbb F_q$-vector-space structure of $\mathbb F_q^{nr}$, since the codes we study are only $ \mathbb{F}_q $-linear.

\begin{definition}We define the inner product between $ \mathbf{c} , \mathbf{d} \in \mathbb{F}_q^{nr} $ as $ \mathbf{c} \cdot \mathbf{d} = c_1d_1 + \cdots + c_{rn} d_{rn} $, where $ \mathbf{c} = (c_1, \ldots, c_{rn}) $ and $ \mathbf{d} = (d_1, \ldots, d_{rn}) $. Given an $ \mathbb{F}_q $-linear code $\mathcal{C}\subseteq \mathbb{F}_q^{nr}$, we define its dual as 
$$\mathcal{C}^\perp = \{ \mathbf{d} \in \mathbb{F}_q^{nr} : \mathbf{c} \cdot \mathbf{d} = 0, \textrm{ for all } \mathbf{c} \in \mathcal{C} \} .$$
\end{definition}

We will use the following three forms of the Singleton bound. The first is the usual Singleton inequality written over the block alphabet, whereas the other two are the corresponding integral relaxations adapted to an $\mathbb F_q$-dimension $k$ that need not be divisible by $r$.

\begin{proposition}[\textbf{\cite{martinez2025linear}}]\label{singleton}
Let $ \mathcal{C} $ be a code of type $ [n,r,k,d] $ and let $ \mathcal{C}^\perp $ be its dual, of type $ [n,r,rn-k,d^\perp] $. Then 
\begin{enumerate}
\item \label{one} $k\leq r(n - d +1)$, 
\item \label{two} $d\leq n-\lceil \frac{k}{r} \rceil+1=n-\lfloor \frac{k-1}{r} \rfloor$, and
\item \label{three} $d^\perp\leq \lfloor \frac{k}{r} \rfloor+1 = \lceil \frac{k+1}{r} \rceil $.
\end{enumerate}
\end{proposition}

As usual, the code is MDS if it attains the bound in Item 1. If $ r \mid k $ (necessary for the code to be MDS), then Items 1 and 2 coincide. However, when $ r \nmid k $, the second bound may be attained but the first one cannot. This motivates the following definition, which slightly extends the usual MDS terminology, since we measure dimension over the subfield $\mathbb F_q$ rather than over $\mathbb{F}_{q^r}$, which is the usual alphabet.

\begin{definition}[\textbf{\cite{martinez2025linear}}]
We say that a code of type $[n,r,k,d]$ is quasi-MDS or QMDS if $d=n-\lceil k/r\rceil+1=n-\lfloor (k-1)/r\rfloor$. A linear code is dually QMDS if both the code and its dual are QMDS. A QMDS code such that $r\mid k$ is an MDS code.
\end{definition}

In \cite[Ex. 12]{martinez2025linear} it is shown that the dual of a QMDS code  is not necessarily also QMDS, which motivates the dually QMDS definition. As shown in \cite[Remark 10]{martinez2025linear}, QMDS codes lie in between MDS and almost MDS codes \cite{faldum} in terms of their size (or dimension over $ \mathbb{F}_q $) relative to their minimum distance. It was also shown in \cite[Th. 4]{martinez2025linear} that the weight distribution of dually QMDS codes only depends on its parameters $ n,r,k,d $, but this is not the case for other QMDS codes.

\section{Polynomial ideal codes} \label{sec PI codes}

Polynomial ideal (PI) codes, introduced in \cite{bhandari2023ideal}, are a general family of $ \mathbb{F}_q $-linear codes in $ \mathbb{F}_q^{nr} $ that recover as particular cases classical Reed--Solomon codes \cite{reed-solomon} and efficient list-decodable codes such as folded Reed--Solomon codes \cite{guruswami2008explicit} and multiplicity codes \cite{rosenbloom}. In this section, we study their maximum possible length, explicitly compute their duals, prove that they are dually QMDS and give their weight distribution. Finally, we extend this code family by considering arbitrary $ P_1, \ldots, P_n $ and study when they are QMDS and/or dually QMDS.

From this section on, we indistinctively write polynomials either as $P(X)$ or simply as $P$.
\begin{definition} \label{def varphi}
Let $P_1,\ldots,P_n\in\mathbb F_q[X]$ be polynomials of degree $r$. For each $i\in[n]$, we identify the quotient $\mathbb F_q[X]/(P_i)$ with $\mathbb F_q^r$ through the ordered basis $(1,X,\ldots,X^{r-1})$ and define
\[
\varphi_i:\mathbb F_q[X]\longrightarrow\mathbb F_q^{1\times r}
\]
as the map that sends $ P(x) $ to its coordinate vector modulo $ P_i(X)$. In other words, 
\[
\varphi_i(P)=\big(\varphi_{i,0}(P),\ldots,\varphi_{i,r-1}(P)\big) \in \mathbb{F}_q^r ,
\]
where $ P \equiv \varphi_{i,0}(P) + \cdots + \varphi_{i,r-1}(P) X^{r-1} \bmod P_i(X) $. Finally, define
\[
\varphi:\mathbb F_q[X]\longrightarrow\mathbb F_q^{nr},
\qquad \textrm{where} \qquad
\varphi(P)=\big(\varphi_1(P),\ldots,\varphi_n(P)\big),
\]
for $ P \in \mathbb{F}_q[X] $.
\end{definition}

\begin{definition}[\textbf{Polynomial ideal codes} \cite{bhandari2023ideal}]\label{def pi codes}
Let $ P_1,\ldots, P_n \in \mathbb{F}_q[X] $ be pairwise coprime polynomials of degree $ r $. For $ k \in [0,nr] $, we define the polynomial ideal (PI) code
$$ \mathrm{PI}_k(P_1, \ldots, P_n) = \{ \varphi(P) : P \in \mathbb{F}_q[X], \deg(P) < k \}\subseteq\mathbb F_q^{nr}. $$
(By convention, $ P = 0 $ is the only polynomial of degree $ < 0 $.)
\end{definition} 

Their generalization was introduced in \cite{wooters}.
\begin{definition} [\cite{wooters}] \label{def GPI}
Let $ P_1,\ldots, P_n \in \mathbb{F}_q[X] $ be pairwise coprime polynomials of degree $ r $. Given $ k \in [0,nr] $ and $ A_1, \ldots, A_n \in {\rm GL}_r(\mathbb{F}_q) $, we define the generalized polynomial ideal (GPI) code
$$ \mathrm{GPI}_k(P_1, \ldots, P_n; A_1, \ldots, A_n) = \{ (\mathbf{c}_1A_1, \ldots, \mathbf{c}_nA_n) : (\mathbf{c}_1, \ldots, \mathbf{c}_n) \in \mathrm{PI}_k(P_1, \ldots, P_n) \}. $$
\end{definition}

Thus, GPI codes are precisely those linear codes that are equivalent to PI codes. In \cite[Th. 2]{martinez2025linear}, we showed that GPI codes are QMDS.

\begin{proposition} [\cite{martinez2025linear}] \label{PI is QMDS} Let $ P_1,\ldots, P_n \in \mathbb{F}_q[X] $ be pairwise coprime polynomials of degree $ r $ and let $ A_1, \ldots, A_n \in {\rm GL}_r(\mathbb{F}_q) $. For $ k \in [0,nr] $, the code $\mathrm{GPI}_k(P_1, \ldots, P_n; A_1, \ldots, A_n) $ from Definition \ref{def GPI} is QMDS of type $ [n,r,k,d] $.
\end{proposition}

To the best of our knowledge, it was not previously known whether PI codes are dually QMDS (except for particular cases such as multiplicity codes \cite[Cor. 31]{martinez2025linear}). In the following sections we will see this and other structural properties of PI codes.

\subsection{Maximum length of PI codes} \label{subsec max length}

Since a PI code is specified by a set of pairwise coprime degree-$r$ polynomials, its maximum length is the maximum cardinality of such a set. We first estimate the number of monic irreducible polynomials of each degree, which will be the basic input for the exact counting argument.

\begin{definition}
    Let $\mu:\mathbb{N}\to\mathbb{Z}$ be the Möbius function:
$$\mu(d)=\left\{ \begin{array}{lcl} 1 & \mbox{if} & d=1, \\ \\ (-1)^k & \mbox{if} & d \mbox{ is the product of }k \mbox{ different primes,} \\ \\ 0 & \mbox{if} & d\mbox{ is divisible by the square of a prime.} \end{array} \right.$$
\end{definition}   

The following result is \cite[Th. 3.25]{lidl}.

\begin{theorem}
The number of monic irreducible polynomials in $\mathbb{F}_q[X]$ of degree $r$ is
$$N_q(r)=\frac{1}{r}\sum_{d \mid  r}\mu(d)q^{r/d}.$$ 
\end{theorem}

We have the following estimates on $ N_q(r) $.

\begin{proposition} \label{bound} The following inequalities hold: 
$$ \begin{array}{lcl} \frac{1}{r}(q^r-q^{r/2})\leq N_q(r)\leq\frac{1}{r}(q^r-q) & \mbox{if} & 2\leq r\leq 5, \\ \\ \frac{1}{r}(q^r-q^{r/2}-q^{r/3})\leq N_q(r)\leq\frac{1}{r}(q^r-q) & \mbox{if} & 6\leq r\leq 29, \\ \\ \frac{1}{r}(q^r-q^{r/2}-q^{r/3}-q^{r/5})\leq N_q(r)\leq\frac{1}{r}(q^r-q) & \mbox{if} & 30\leq r. \end{array} $$
\end{proposition}
\begin{proof}
    Given $r=p_1^{e_1}\cdots p_n ^{e_n}$, where $p_1<\ldots<p_n$ are primes and $e_1,\ldots,e_n$ are positive integers, if $n=1$, then $N_q(r)=\frac{1}{r}\left(q^r-q^{r/p_1}\right)$
and the desired lower bound follows immediately. Hence, assume that $n\geq2$. We can rewrite the formula for $N_q(r)$ as a sum of powers arranged in descending order:
$$N_q(r)=\frac{1}{r}\left(q^r-q^{\frac{r}{p_1}} -q^{\frac{r}{p_2}}-\cdots-q^{\frac{r}{p_i}}+q^{\frac{r}{p_1p_2}}+\cdots\right),$$
where $p_i$ is the largest prime such that $p_i<p_1p_2$. By definition of $\mu(d)$, each exponent appearing in the sum is associated with a squarefree divisor of $r$, and hence no exponent is repeated.  Also $q^r>\sum_{i=0}^{r-1}q^j$, thus the term with power $\frac{r}{p_1p_2}$ is greater than all the following terms of the sum combined. This yields:
$$N_q(r)\geq\frac{1}{r}\left(q^r-q^{\frac{r}{p_1}} -q^{\frac{r}{p_2}}-\dots-q^{\frac{r}{p_i}}\right).$$
Now, if $r$ is not divisible by $2$, then $2<p_1$ and $q^{\frac{r}{2}}>\displaystyle\sum_{j=0}^{\left\lfloor\frac{r}{2}\right\rfloor-1}q^j$, thus: 
$$N_q(r)\geq\frac{1}{r}\left(q^r-q^{\frac{r}{2}}+q^{\frac{r}{2}}-q^{\frac{r}{p_1}} -\dots-q^{\frac{r}{p_i}}\right)\geq \frac{1}{r}\left(q^r-q^{\frac{r}{2}}\right).$$
Next, if $r$ is divisible by $2$ but not divisible by $3$, then similarly,
$$N_q(r)\geq\frac{1}{r}\left(q^r-q^{\frac{r}{2}}-q^{\frac{r}{3}}+q^{\frac{r}{3}}-q^{\frac{r}{p_2}}-\dots-q^{\frac{r}{p_i}}\right)\geq \frac{1}{r}\left(q^r-q^{\frac{r}{2}}-q^{\frac{r}{3}}\right).$$   
 We note that the previous inequality also holds when $p_i=2$. Analogously, when $r$ is divisible by $2$ and $3$ but not divisible by $5$ we can deduce that $N_q(r)\geq \frac{1}{r}(q^r-q^{\frac{r}{2}}-q^{\frac{r}{3}}-q^{\frac{r}{5}})$. Now, when $r$ is divisible by $2$, $3$ and $5$ we see that $p_1p_2=6$ and $p_i=5$ and we conclude similarly that $N_q(r)\geq\frac{1}{r}(q^r-q^{\frac{r}{2}}-q^{\frac{r}{3}}-q^{\frac{r}{5}})$.  For the upper bound we see that $rN_q(r)-q^r$ is a negative number divisible by $q$, hence 
 $$N_q(r)\leq \frac{q^r-q}{r}.$$
\end{proof}

The preceding estimates immediately give the asymptotic behaviour of $N_q(r)$.

\begin{corollary}\label{cor:asymptotically}
For fixed $q$ and $ r $, respectively,
\[
\lim_{r\to\infty}\frac{N_q(r)}{q^r/r}=1 \quad \textrm{and} \quad \lim_{q\to\infty}\frac{N_q(r)}{q^r/r}=1.
\]
\end{corollary}

We next show the monotonicity of $ N_q(r) $ in $ r $, which will be used to maximize the number of pairwise coprime degree-$r$ polynomials.

\begin{theorem}\label{irre} For all $q\geq2$ we have that
    $N_q(r-1)<N_q(r)$, except the cases $N_2(1)>N_2(2)$ and $N_3(1)=N_3(2)$.    
\end{theorem}
\begin{proof}
We assume $ r \geq 4 $, since the other cases can be checked directly. By the bounds from Proposition \ref{bound}, it is enough to prove that
  
    \begin{align*}
    &(r-1)\left(q^r-q^{r/2}-q^{r/3}-q^{r/5}\right)-r(q^{r-1}-q)>0.
    \end{align*}
Let us assume that $q\geq2$ and $r\geq6$. This implies that $r-2>\frac{r}{2}$ thus $q^r,q^{r-1},q^{r-2},q^{r/2},q^{r/3}$ and $q^{r/5}$ are all different powers. Using that $q^r-1=(q-1)\sum_{i=0}^{r-1}q^i$,  if  $r-1>q$ we see that:
 \begin{align*}
    &(r-1)\left(q^r-q^{r/2}-q^{r/3}-q^{r/5}\right)-r(q^{r-1}-q)>\\
    &(r-1)\left(q^r-q^{r-1}-q^{r/2}-q^{r/3}-q^{r/5}\right)-q^{r-1}>\\
    & (r-1)q^{r-2}-q^{r-1}>0.
    \end{align*}
Similarly, if $q>r-1$ we see that: 
 \begin{align*}
    &(r-1)\left(q^r-q^{r/2}-q^{r/3}-q^{r/5}\right)-r(q^{r-1}-q)=\\
    &q^{r-1}\left((r-1)q-r\right)-(r-1)\left(q^{r/2}+q^{r/3}+q^{r/5}\right)+rq=\\
    &q^{r-1}\left((r-1)(q-1)-1\right)-(r-1)\left(q^{r/2}+q^{r/3}+q^{r/5}\right)+rq>\\
    &q^{r-1}-(r-1)q^{r-2}>0.
    \end{align*}

 The first equality holds thanks to $r-2>\frac{r}{2}$ ensuring that the fractional powers do not annihilate  $q^{r-2}$. The second equality is obtained factoring $q^{r-1}$. For the third inequality we use that $r-2>\frac{r}{2}$ once again. Finally, it is easy to check the case $q=r-1$ using a similar reasoning and we can simply compute the cases $r=3, 4$ and $5$. 
 
    \end{proof}
    
We can now express the maximum length of PI codes in terms of the numbers $N_q(r)$. Recall from Definition \ref{def pi codes} that the maximum length $ n $ of the PI code $ \mathrm{PI}_k(P_1, \ldots, P_n) $ in $ \mathbb{F}_q^{nr} $ is $ n = M_q(r) $, the maximum possible size of a set $ \{ P_1, P_2, \ldots, P_n \} $ of pairwise coprime degree-$ r $ polynomials in $ \mathbb{F}_q[X] $.

\begin{theorem}\label{max}
For $r>3$, the maximum cardinality $M_q(r)$ of a set of pairwise coprime polynomials of degree $r$ in $\mathbb{F}_q[X]$ is
$$M_q(r)= N_q(1)+N_q(2)+\cdots+\,N_q\left(\left\lfloor\frac{r}{2}\right\rfloor\right)+N_q(r).
$$
\end{theorem}
\begin{proof}
Let $\mathcal F$ be a family of pairwise coprime degree-$r$ polynomials in $ \mathbb{F}_q[X] $. The irreducible members of $\mathcal F$ contribute at most $N_q(r)$ elements. Every reducible polynomial of degree $r$ has an irreducible factor of degree at most $\lfloor r/2\rfloor$. Since the members of $\mathcal F$ are pairwise coprime, distinct reducible members must contain distinct such irreducible factors. Consequently,
\[
|\mathcal F|\leq N_q(r)+\sum_{i=1}^{\lfloor r/2\rfloor}N_q(i).
\]

It remains to attain this bound. Include first all monic irreducible polynomials of degree $r$. For each $1\leq i<r/2$, pair every monic irreducible polynomial $Q$ of degree $i$ with a distinct monic irreducible polynomial $R_Q$ of degree $r-i$, and include the product $QR_Q$. Such a choice is possible because $N_q(r-i)\geq N_q(i)$ for $r>3$, by Theorem \ref{irre}. The sets of complementary factors $ R_Q $ used for different $i = \deg(Q)$ have different degrees and are therefore disjoint. If $r$ is even, for each irreducible $Q$ of degree $r/2$ include $Q^2$. The resulting degree-$r$ polynomials are pairwise coprime and are also coprime to all irreducible polynomials of degree $r$. Hence the upper bound is attained.
\end{proof}

As expected, $ M_q(r) $ (i.e., the maximum length of PI codes) is also strictly increasing in $ r $.

\begin{corollary}
For all $q\geq2$ and all $r\geq2$ we have that
    $M_q(r-1)<M_q(r)$.    
\end{corollary}
\begin{proof}
First, we see that $M_q(1)=N_q(1)<M_q(2)=N_q(1)+N_q(2)<M_q(3)=N_q(1)+N_q(3)$. 
Now, for $r\geq4$, thanks to Theorem \ref{max} and Theorem
\ref{irre} we see that, if $r$ is odd,
$$
M_q(r)-M_q(r-1)=N_q(r)-N_q(r-1)>0,
$$
whereas, if $r$ is even,
$$
M_q(r)-M_q(r-1)
=
N_q\left(\frac r2\right)+N_q(r)-N_q(r-1)>0.
$$
\end{proof}
The previous results also give an asymptotic estimate for the maximum length of PI codes.

\begin{corollary}\label{cor:asymptotic-Mq}
For fixed $q \geq 2$ and $ r \geq 4 $, respectively,
\[
\lim_{r\to\infty}\frac{M_q(r)}{q^r/r}=1 \quad \textrm{and} \quad \lim_{q\to\infty}\frac{M_q(r)}{q^r/r}=1.
\]
\end{corollary}
\begin{proof}
Defining $ t = \lfloor r/2 \rfloor $, since $ N_q(i) \leq q^i $, we have that
$$ N_q(r) \leq M_q(r) = N_q(r) + \sum_{i=1}^t N_q(i) \leq N_q(r) + t q^t. $$
Dividing by $ q^r/r $, we obtain
$$ \frac{N_q(r)}{q^r/r} \leq \frac{M_q(r)}{q^r/r} \leq \frac{N_q(r)}{q^r/r} + rt q^{t-r}, $$
and the result follows by the Sandwhich criterion and Corollary \ref{cor:asymptotically}.
\end{proof}


\subsection{Duality and weight distribution of PI codes}\label{sec:PI-duality}
We now compute the dual of a general polynomial ideal code with respect to the standard inner product of $\mathbb F_q^{nr}$. In \cite{wooters}, the dual is computed using matrices over the algebraic closure of $\mathbb F_q$. Our result shows that all block matrices may be chosen over $\mathbb F_q$ itself. In other words, we compute the dual over $ \mathbb{F}_q $ and not only over its algebraic closure. Throughout this subsection, unless stated otherwise, $P_1,\ldots,P_n\in\mathbb F_q[X]$ are pairwise coprime polynomials of degree $r$. 

\begin{definition} Given $P=a_0+\cdots+a_{r}X^{r}\in\mathbb{F}_q[X]$  such that $a_r\neq0$ we say that the following matrix $M\in\mathbb{F}_q^{r\times r}$ is the companion matrix of $P$:
 \[
M=\begin{pmatrix}
0 & 1 & 0 & \cdots & 0\\
0 & 0 & 1 & \cdots & 0\\
\vdots & \vdots& \vdots& \ddots & \vdots\\
0 & 0 & 0 & \cdots & 1\\
- a_0a_r^{-1} & - a_1a_r^{-1} & - a_2a_r^{-1} & \cdots & - a_{r-1}a_r^{-1}
\end{pmatrix}.
\]  
\end{definition}

\begin{lemma} \label{companion}Let $M_i\in\mathbb{F}_q^{r\times r}$ be the companion matrix of $P_i\in\mathbb{F}_q[X]$ with $\mathrm{deg}(P_i)= r$. Then, given $F\in\mathbb{F}_q[X]$  with $\mathrm{deg}(F)< r$, we have that $\varphi_i(XF(X))=\varphi_i(F(X))M_i$. In other words, the coordinates of $\varphi_i(F)M_i\in\mathbb{F}_q^r$  correspond to the coefficients of $XF(X)\bmod P_i$.
    
\end{lemma}

\begin{proof}
    Let $P_i=a_0+\cdots+a_{r}X^{r}\in\mathbb{F}_q[X]$  and $F=f_0+\cdots+f_{r-1}X^{r-1}\in\mathbb{F}_q[X]$. We see that each row $j\in[r]$ of the companion matrix $M_i$ is the vector whose coordinates are the coefficients of $XX^{j-1}=X^j\bmod{P_i}$ because $a_rX^r\equiv -a_0-\cdots-a_{r-1}X^{r-1}\bmod{P_i}$. We conclude by linearity of $\varphi_i$.
\end{proof}

The following lemma is the blockwise analogue of Lagrange interpolation, with congruence classes modulo the $P_i$ replacing evaluations at field elements.

\begin{lemma} \label{interpolacion por bloques}
There exist unique polynomials $S_{ij}(X)\in\mathbb F_q[X]$, for $i\in[n]$ and $j\in[0,r-1]$, with $\deg(S_{ij})<nr$ such that every $H(X)\in\mathbb{F}_q[X]$ with $\deg(H)<nr$ satisfies
\begin{equation}
H(X)= \sum_{i=1}^n\sum_{j=0}^{r-1} \varphi_{ij}(H)\,S_{i,j}(X).
\label{eq H(X)}
\end{equation}
\end{lemma}

\begin{proof}
 For each $i\in[n]$ we can define
\[
N_i(X):=\prod_{j\neq i}P_j(X),\mbox{ and let }  T_i(X)\equiv N_i(X)^{-1}\bmod{P_i(X)}.
\]
If $E(X)=\prod_{i=1}^n P_i(X)$, then we can choose $ S_i(X)\equiv N_i(X)T_i(X)\bmod{E(X)}$  with $S_i(X)$ such that $\deg (S_i)<nr$. Then we see that $S_i(X)\equiv 1\ \mathrm{mod}\ P_i(X)$ and $S_i(X)\equiv 0\ \mathrm{mod}\ P_{\ell}(X)$ for $\ell\ne i$. For $j\in[0,r-1]$ set
$S_{i,j}(X)\equiv X^{j} S_i(X)\bmod E(X)$ such that  $\deg(S_{i,j})<nr$. We see that, modulo $P_i(X)$, the element $S_{i,j}(X)$ reduces to $X^{j}$, while $S_{\ell,j}\equiv 0\ \mathrm{mod}\ P_i$ for $\ell\neq i$.
Thus the right-hand side of (\ref{eq H(X)}) is congruent to $H(X)$ modulo every $P_i(X)$ and hence modulo $E(X)$ by the Chinese remainder theorem. Since both sides have degree $<nr=\deg(E)$, they are equal in $\mathbb F_q[X]$. 

For uniqueness, the residue conditions
\[
S_{ij}\equiv X^j\pmod{P_i} \quad \textrm{and} \quad
S_{ij}\equiv0\pmod{P_\ell}, \textrm{ for }\ell\neq i,
\]
imply that $ S_{ij} $ is unique modulo $E$ by the Chinese remainder theorem. Finally, by the degree restriction $\deg(S_{ij})<nr$, we have that $ S_{ij} $ is unique in $ \mathbb{F}_q[X] $.
\end{proof}

We now need some technical lemmas.

\begin{lemma}\label{varphi del producto}
 Given $P,Q \in \mathbb{F}_q[X]$ we have that 
$$\varphi_i(PQ)=\varphi_i(P)\left(\sum_{j=0}^{r-1}\varphi_{ij}(Q)M_i^{j}\right),$$
where $M_i$ is the companion matrix of $P_i$.
\end{lemma}

\begin{proof}
    In the following equations, the first equality holds because $\varphi_i$ only depends on the class modulo $P_i$ of its input, the second equality holds because $\varphi_i$ is linear and the last equality holds thanks to the main property of the companion matrix $M_i$, explained in Lemma \ref{companion}:    $$\varphi_i(PQ)=\varphi_i\left(P(X)\sum_{j=0}^{r-1}\varphi_{ij}(Q)X^j\right)=\sum_{j=0}^{r-1}\varphi_{ij}(Q)\varphi_i\left(X^jP(X)\right)=\sum_{j=0}^{r-1}\varphi_{ij}(Q)\varphi_i(P)M_i^j.$$
\end{proof}

For a polynomial $F(X)=\sum_j f_jX^j \in \mathbb{F}_q[X] $, we write $\operatorname{coef}(F,j)=f_j \in \mathbb{F}_q $.

\begin{lemma}\label{A_i invertible}
    Let $A_1,\ldots,A_n\in\mathbb{F}_q^{r\times r}$ be such that, for any given  $P,Q\in\mathbb{F}_q[X]$ with $\deg(PQ)<nr$ we have that
    $$\mathrm{coef}(PQ,nr-1)=\sum_{i=1}^n\varphi_i(P)A_i\varphi_i(Q)^{\top}.$$
    Then each $A_i$ is invertible. 
\end{lemma}
 
\begin{proof}
    If we suppose that for some $i\in[n]$ we have that $A_i$ is not invertible,  then there exists some non zero $\mathbf{u}_i\in\mathbb{F}_q^r$ such that $\mathbf{u}_iA_i= \mathbf{0} $. Let us consider $\mathbf{u}=(\mathbf{u}_j)_{j=1}^n\in\mathbb{F}_q^{nr}$ with $\mathbf{u}_j= \mathbf{0} \in\mathbb{F}_q^r$  for $j\neq i$. By the Chinese remainder theorem, the restriction of $\varphi$ to the polynomials of degree $<nr$ is an $\mathbb F_q$-linear isomorphism onto $\mathbb F_q^{nr}$. Hence there exists a nonzero polynomial $P\in\mathbb F_q[X]$ with $\deg(P)<nr$ such that $\varphi(P)=\mathbf u$. Thus by how have we constructed $\mathbf{u}$, we see that  $\varphi(P)\mathrm{diag}(A_1,\ldots,A_n)= \mathbf{0} $ which implies that $\mathrm{coef}(PQ,nr-1)=\varphi(P)\mathrm{diag}(A_1,\ldots,A_n)\varphi(Q)^{\top}=0$, for all $Q\in\mathbb{F}_q[X]$ such that $\mathrm{deg}(PQ)<nr$. This is absurd because if we choose $Q(X)=X^{nr-1-\mathrm{deg}(P)}$ we see that $\mathrm{deg}(PQ)<nr$ but, however,  $\mathrm{coef}(PQ,nr-1)=\mathrm{coef}(P,\mathrm{deg}(P)) \neq0$.
\end{proof}

Finally, we show that the dual of a PI code is a GPI code over a finite field (and not only over its algebraic closure as in \cite{wooters}).

\begin{theorem}\label{thm:PIC-dual}
Given $\mathcal{C}=\mathrm{PI}_k(P_1,\dots,P_n)\subseteq \mathbb{F}_q^{nr}$, let $M_i$ be the companion matrix of each $P_i$. Then there exist $\mathbf{a}_1,\ldots,\mathbf{a}_n\in\mathbb{F}_q^r$ depending only on $P_1,\ldots,P_n$ such that:
$$\mathcal{C}^{\perp}=\mathrm{GPI}_{nr-k}(P_1, \ldots, P_n; A_1, \ldots, A_n),$$
where $A_i=\big(M_i^{j-1}\mathbf{a}_i^{\top}\big)_{j=1}^{r}\in\mathbb{F}_q^{r\times r}$  are invertible matrices.

\end{theorem}

\begin{proof}
Let $P,Q\in\mathbb{F}_q[X]$ satisfying that $\deg(PQ)<nr$. If we use Lemma \ref{interpolacion por bloques} in order to compute $c_{nr-1}=\mathrm{coef}(PQ,nr-1)$ we see that
$$c_{nr-1}=\sum_{i=1}^n\sum_{j=0}^{r-1}\varphi_{ij}(PQ)\mathrm{coef}(S_{ij},nr-1)=\sum_{i=1}^n\varphi_i(PQ)\mathbf{a}_i^{\top},$$
where $\mathbf{a}_i=\left(\mathrm{coef}(S_{ij},nr-1)\right)_{j=0}^{r-1} \in \mathbb{F}_q^r $.
Now we will use Lemma \ref{varphi del producto} with the goal of splitting $\varphi_i(PQ)$ into an expression depending on $\varphi_i(P)$ and $\varphi_i(Q)$.
\begin{equation*}
\begin{split}
c_{nr-1} & =\sum_{i=1}^n\varphi_i(PQ)\mathbf{a}_i^{\top}=\sum_{i=1}^n\varphi_i(P)\left(\sum_{j=0}^{r-1}\varphi_{ij}(Q)M_i^j\right)\mathbf{a}_i^{\top} \\
 & =\sum_{i=1}^n\varphi_i(P)A_i\varphi_i(Q)^{\top}=\varphi(P)\mathrm{diag}(A_1,\ldots,A_n)\varphi(Q)^{\top},
\end{split}
\end{equation*}
where $A_i=\big(M_i^{j-1}\mathbf{a}_i^{\top}\big)_{j=1}^{r}\in\mathbb{F}_q^{r\times r}$.
Lemma \ref{A_i invertible} shows that each $A_i\in\mathbb F_q^{r\times r}$ is invertible. Now let $Q\in\mathbb F_q[X]$ satisfy $\deg(Q)<k$ and let $P\in\mathbb F_q[X]$ satisfy $\deg(P)<nr-k$. Then $\deg(PQ)<nr-1$. Hence $0=c_{nr-1}=\varphi(P)\mathrm{diag}(A_1,\ldots,A_n)\varphi(Q)^{\top}$. In other words, every vector in $\mathrm{PI}_{nr-k}(P_1,\dots,P_n)\cdot \mathrm{diag}(A_1,\dots,A_n)$
is orthogonal to every codeword in $\mathcal{C}$, so
\[
\mathrm{GPI}_{nr-k}(P_1, \ldots, P_n; A_1, \ldots, A_n)\subseteq \mathcal{C}^{\perp}.
\]
 Finally the inclusion is actually an equality because both spaces have $\mathbb{F}_q$-dimension $nr-k$, as we wanted to prove.
\end{proof}

Therefore, we deduce that PI codes (and therefore GPI codes) are dually QMDS.
%

\begin{theorem}\label{PI dually QMDS}
Every polynomial ideal code $\mathcal C=\mathrm{PI}_k(P_1,\ldots,P_n)\subseteq\mathbb F_q^{nr}$ is dually QMDS.
\end{theorem}

\begin{proof} $\mathcal{C}$ is QMDS by Proposition \ref{PI is QMDS}. Now by Theorem \ref{thm:PIC-dual} and Proposition \ref{prop isometries}, the dual $\mathcal{C}^{\perp}$ is equivalent to $\mathrm{PI}_{nr-k}(P_1, \ldots, P_n)$, which is again QMDS, and we are done. 
\end{proof}

We computed the folded weight distribution of a dually QMDS code in \cite[Th. 4]{martinez2025linear}, which is determined by its parameters $ n,r,k,d $. Combined with Theorem \ref{PI dually QMDS}, we obtain the following.

\begin{theorem} 
Let $\mathcal C=\mathrm{PI}_k(P_1,\ldots,P_n)\subseteq\mathbb F_q^{nr}$, and let $(W_j)_{j=0}^n$ be its folded weight distribution. Then $W_0=1$, $W_j=0$ for all $j\in[d-1]$, and if $ j \in [d,n] $, then
$$ W_j = \binom{n}{j} \sum_{i=0}^{j - d}(-1)^i \binom{j}{i} \big(q^{k-r(n-j+i)}-1\big),\qquad \textrm{where } d=n-\left\lceil\frac{k}{r}\right\rceil+1. $$
\end{theorem}

\begin{remark}
All the preceding properties are invariant under linear isometry. In particular, every GPI code is dually QMDS, has the same weight distribution as the corresponding PI code, and its dual is equivalent to a PI code of complementary dimension.
\end{remark}

\subsection{Polynomial residue codes: Extending PI codes} \label{subsec PR codes}

We now introduce a class of codes that generalize PI codes (Definition \ref{def pi codes}).

\begin{definition}[\textbf{Polynomial residue codes}]\label{def pr codes}
Let $ P_1,\ldots, P_n \in \mathbb{F}_q[X] $ be arbitrary nonzero polynomials of degree at most $ r $. For $ k \in [0,nr] $, we define the polynomial residue (PR) code
$$ \mathrm{PR}_k(P_1, \ldots, P_n) = \{ \varphi(P) : P \in \mathbb{F}_q[X], \deg(P) < k \}\subseteq\mathbb F_q^{nr}. $$
Here, $ \varphi : \mathbb{F}_q[X] \longrightarrow \mathbb{F}_q^{rn} $ is as in Definition \ref{def varphi}, but adding $ r-d_i $ zeros at the end of $ \varphi_i(P) \in \mathbb{F}_q^r $, for $ P \in \mathbb{F}_q[X] $, if $ d_i = \deg(P_i) < r $, for $ i \in [n] $. 
\end{definition} 

In other words, PI codes are PR codes with pairwise coprime polynomials $ P_1, \ldots, P_n $, all of degree exactly $ r $.

Throughout this subsection we fix arbitrary $ k \in [0,rn] $, monic nonzero $ P_1, \ldots, P_n $ of degree at most $ r $, and the corresponding PR code $ \mathcal{C} = \mathrm{PR}_k(P_1, \ldots,P_n) $. Notice that assuming that $ P_1,\ldots,P_n $ are monic does not change the code $ \mathcal{C} $. Throughout this subsection, we also define
$$ d_i = \deg(P_i) \qquad \textrm{and} \qquad L_J = {\rm lcm}\{ P_j : j \in J \}, $$
for $ i \in [n] $ and $ J \subseteq [n] $, where we define $ L_\varnothing = 1 $ and $ L = L_{[n]} $. We start by computing the dimensions of PR codes.

\begin{proposition} \label{prop dim PR codes}
We have that $ \dim(\mathcal{C}) = \min \{ k, \deg(L) \} $. 
\end{proposition} 
\begin{proof}
The kernel of the $ \mathbb{F}_q $-linear map $ \varphi : \mathbb{F}_q[X] \longrightarrow \mathbb{F}_q^{rn} $ is given by the polynomials $ P \in \mathbb{F}_q[X] $ divisible by $ L $. If we restrict the domain to polynomials of degree less than $ k $, such a kernel has dimension $ \max \{ 0,k-\deg(L) \} $. Since $ \mathcal{C} $ is the image of the subspace of polynomials of degree $ < k $ by $ \varphi $, we deduce the formula for its dimension.
\end{proof}

In particular, we may assume that $ k \leq \deg(L) $ for the rest of the subsection and assume, without loss of generality, that it is the dimension of $ \mathcal{C} $.
 
We will need punctured and shortened codes. The next lemma has the same proof as \cite[Th. 1.5.7]{pless}. 

\begin{lemma} \label{lemma punctured shortened folded}
For $ J \subseteq [n] $, define $ \pi_J : \mathbb{F}_q^{rn} \longrightarrow \mathbb{F}_q^{|J|r} $ by $ \pi_J(\mathbf{c}_1, \ldots, \mathbf{c}_n) = (\mathbf{c}_j)_{j \in J} $, for $ \mathbf{c}_1, \ldots, \mathbf{c}_n $ $ \in \mathbb{F}_q^{r} $. For an $ \mathbb{F}_q $-linear code $ \mathcal{D} \subseteq \mathbb{F}_q^{rn} $, define the punctured and shortened codes of $ \mathcal{D} $ at $ J $, respectively, as
$$ \mathcal{D}^J = \pi_J(\mathcal{D}) \qquad \textrm{and} \qquad \mathcal{D}_J = \pi_J(\mathcal{D} \cap \ker(\pi_{[n] \setminus J})). $$
Then we have that $ (\mathcal{D}^J)^\perp = (\mathcal{D}^\perp)_J $ and $ (\mathcal{D}_J)^\perp = (\mathcal{D}^\perp)^J $.
\end{lemma}

\begin{lemma}\label{lem:ars-projections folded}
For $J\subseteq[n]$, $ \dim(\mathcal{C}^J)=\min\{k,\deg (L_J)\} $ and $ \dim((\mathcal{C}^\perp)_J)= |J|r - \min\{k,\deg (L_J)\} $.
\end{lemma}
\begin{proof}
Notice that $ \mathcal{C}^J $ is the PR code with degree restriction $ k $ and polynomials $ P_j $, for $ j \in J $. Thus by Proposition \ref{prop dim PR codes}, $ \dim(\mathcal{C}^J) = \min\{k,\deg (L_J)\} $. Now, since $ (\mathcal{C}^\perp)_J = (\mathcal{C}^J)^\perp $ by Lemma \ref{lemma punctured shortened folded}, we deduce the formula for $ \dim((\mathcal{C}^\perp)_J) $. 
\end{proof}

We may now characterize when PR codes are QMDS. This characterization depends on the polynomials $ L_J $ instead of the parameters of the code, but it will be enough to show that QMDS PR codes cannot be ``too long''.

\begin{proposition}\label{prop:QMDS PR}
If $ k > 0 $ and $s=\lceil k/r\rceil$, then $ \C $ is QMDS if and only if $ \deg(L_J) \geq k $, for all $ J \subseteq [n] $ such that $ |J| = s $.
\end{proposition}
\begin{proof}
There exists a nonzero codeword in $ \C $ that vanishes in $ J $ if and only if $ \dim(\C^J) < k $, which by Lemma \ref{lem:ars-projections folded} is equivalent to $ \deg(L_J) < k $. Hence by the Singleton bound (Proposition \ref{singleton}, Item \ref{two}), $ \C $ is QMDS if and only if $ \deg(L_J) \geq k $, for all $ J \subseteq [n] $ such that $ |J| = s $. 
\end{proof}

As a consequence, we show that QMDS PR codes cannot be longer than allowed by the MDS conjecture. 

\begin{corollary}
If $ k > r $ and $ \C $ is QMDS, then $ n \leq q^r $.
\end{corollary}
\begin{proof}
We have $ s = \lceil k/r\rceil \geq 2 $ and $ n \geq s $. If there exist $ i,j \in [n] $ such that $ i \neq j $ and $ P_i $ divides $ P_j $, then for any $ J \subseteq [n] $ with $ i,j \in J $ and $ |J| = s $, we have 
$$ \deg(L_J) = \deg(L_{J \setminus \{i \}}) \leq (s-1)r < k. $$
Thus $ \C $ would not be QMDS by Proposition \ref{prop:QMDS PR}. Since it is, we deduce that no two different polynomials in $ P_1, \ldots, P_n $ divide each other. If we set 
$$ Q_i = X^{r-d_i} P_i \in \mathbb{F}_q[X] , $$
for $ i \in [n] $, then $ \deg({\rm lcm}\{ Q_j : j \in J \}) \geq \deg(L_J) $, for all $ J \subseteq [n] $. Thus $ \mathrm{PR}_k(Q_1, \ldots,Q_n) $ has dimension $ k $ by Proposition \ref{prop dim PR codes}, and is also QMDS by Proposition \ref{prop:QMDS PR}. Finally, $ Q_1, \ldots, Q_n $ are distinct. If not, there would be $ i,j \in [n] $ such that $ i \neq j $ and $ P_j = X^{d_j-d_i} P_i $, which contradicts that $ P_i $ and $ P_j $ do not divide each other. Hence $ Q_1, \ldots, Q_n \in \mathbb{F}_q[X] $ are distinct monic polynomials of degree $ r $, and therefore $ n \leq q^r $.
\end{proof}
 
We conclude by showing that, for most parameters, PR codes are dually QMDS (in particular MDS if $ r $ divides $ k $) if and only if they are PI codes.

\begin{theorem}
With the convention $ \dd(\{ \mathbf{0} \}) = n+1 $, we have
\[
\dd(\C^\perp)=
\begin{cases}
1,&\textrm{if }k<r\textrm{ or }d_i<r, \textrm{ for some } i \in [n],\\[2pt]
2,&\textrm{if }k\geq r,\ d_i=r\textrm{ for all }i \in [n] \textrm{ and for some } i \neq j , \gcd(P_i,P_j)\ne1,\\[2pt]
\lfloor k/r\rfloor+1,&\textrm{if }k\geq r,\ d_i=r\textrm{ for all }i \in [n] \textrm{ and }P_1, \ldots, P_n\textrm{ are pairwise coprime}.
\end{cases}
\]
In particular, $\C^\perp$ is QMDS if and only if
\begin{enumerate}
\item $0\leq k<r$; or
\item $r\leq k<2r$ and $ d_i =r$ for all $i\in[n]$; or
\item $2r\leq k\leq nr$ and $P_1, \ldots, P_n$ are pairwise coprime of degree $r$.
\end{enumerate}
\end{theorem} 
\begin{proof}
By Lemma \ref{lem:ars-projections folded}, there exists a nonzero codeword in $\C^\perp$ that vanishes outside of a subset $J \subseteq [n]$ if and only if $\min\{k,\deg(L_J)\}< |J|r$.

Assume first that $k<r$ or that there exists $ i \in [n] $ such that $d_i<r$. Choosing $J=\{i\}$, we deduce that there is a nonzero codeword in $ \C^\perp $ that vanishes outside of $ J $, thus $\dd(\C^\perp)=1$.

Assume now that $k\geq r$ and $d_i=r$, for all $i \in [n]$. We deduce similarly that $\dd(\C^\perp)\geq2$. Now, if $\gcd(P_i,P_j)\ne1$ for some $ i,j \in [n] $ with $i\ne j$, then
$$ \deg(L_{\{i,j\}}) =2r-\deg(\gcd(P_i,P_j))<2r, $$
and therefore $\dd(\C^\perp)=2$.

Finally, if $P_1, \ldots, P_n $ are pairwise coprime, then $ \deg(L_J) = |J|r $ for all $J \subseteq [n] $. Hence $\min\{k,\deg(L_J)\} \geq |J|r$ for all $ J \subseteq [n] $ with $ k \geq |J|r $. In particular, for $ k < rn $, there exists no nonzero codeword in $ \C^\perp $ that vanishes outside of $ J $, for all $ J \subseteq [n] $ such that $ |J| = \lfloor k/r\rfloor $. Thus $ \dd(\C^\perp) = \lfloor k/r\rfloor+1 $ if $ k < rn $. If $ k = rn $, then $ \C^\perp = \{ \mathbf{0} \} $ and $ \dd(\C^\perp) = n+1 $ by convention.
\end{proof}

Thus we conclude with the characterization of dually QMDS PR codes.

\begin{corollary}\label{cor:dually-QMDS PR}
The PR code $\C$ is dually QMDS if and only if 
\begin{enumerate}
\item $0\leq k\leq r$ and $d_i\geq k$ for all $i\in[n]$; or
\item $r<k<2r$, $d_i=r$ for all $i\in[n]$, and for all $ i,j \in [n] $ with $ i \neq j $,
$$ \deg\big(\gcd(P_i,P_j)\big)\leq2r-k ; \textrm{ or} $$
\item $2r\leq k\leq nr$ and $P_1, \ldots, P_n$ are pairwise coprime of degree $r$ (i.e., $ \C $ is a PI code).
\end{enumerate}
\end{corollary}

Notice that for $ k \in [0,r] $ (and thus for $ k \in [(n-1)r,nr] $ by taking duals) we may obtain dually QMDS codes of unrestricted length. This is similar to what happens for $ r=1 $ (i.e., the classical Hamming metric) and the repetition code and its dual.

\section{Additive Reed--Solomon codes} \label{sec aRS}

Additive Reed--Solomon (aRS) codes were first introduced in \cite[Sec. 4]{yadav} and later studied in \cite{bossaller2025nonlinear}. They are a natural extension of classical Reed--Solomon codes \cite{reed-solomon} to $ \mathbb{F}_q $-linear codes in $ \mathbb{F}_{q^r}^n $. In this section, we show that they are a particular case of PI codes in most of the cases when they are dually QMDS. In these cases, we give their duals, weight distributions and permutation group, and show that they asymptotically achieve the maximum length of PI codes. At the end of the section, we will study when arbitrary aRS codes are QMDS and/or dually QMDS.

Given $\alpha\in\mathbb F_{q^r}$, we denote its minimal polynomial over
$\mathbb F_q$ by $\min(\alpha,\mathbb F_q)\in\mathbb F_q[X]$. We refer to the conjugacy class of $\alpha$ as its orbit under the Frobenius map
$x\mapsto x^q$ in $\mathbb F_{q^r}$. In particular, two elements belong to the
same conjugacy class if and only if they have the same minimal polynomial over
$\mathbb F_q$. See \cite[Ch. 1]{lidl}.

\begin{definition}[\cite{yadav}]
Let $\bm\alpha=(\alpha_1,\ldots,\alpha_n)\in\mathbb F_{q^r}^n$ and let $ k \in [0,nr]$. The additive Reed--Solomon (aRS) code with evaluation vector $\bm\alpha$ and degree parameter $k$ is
\[
\mathrm{aRS}_q(\bm\alpha,k)
=
\big\{(P(\alpha_1),\ldots,P(\alpha_n)):
P\in\mathbb F_q[X],\ \deg(P)<k\big\}
\subseteq\mathbb F_{q^r}^n.
\]
\end{definition}

As pointed out in \cite{bossaller2025nonlinear, yadav}, the parameter
$k$ does not necessarily coincide with the $\mathbb F_q$-dimension of the
code. We give a formula for the dimension below.

\begin{proposition}\label{prop:dim-aRS}
Given $\bm\alpha\in\mathbb F_{q^r}^n$, let $L(X)
=
\operatorname{lcm}\{
\min(\alpha_1,\mathbb F_q),\ldots,
\min(\alpha_n,\mathbb F_q)
\}$. Then
\[
\dim_{\mathbb F_q}(\mathrm{aRS}_q(\bm\alpha,k))
=
\min\{k,\deg(L)\}.
\]

\end{proposition}
\begin{proof}
The code is generated over $\mathbb F_q$ by the evaluations of $1,X,\ldots,X^{k-1}$ at $\alpha_1,\ldots,\alpha_n$. Hence these $k$ generators are linearly
dependent if and only if there exists a nonzero polynomial
$P\in\mathbb F_q[X]$ with $\deg(P)<k$ such that $P(\alpha_i)=0$ for all $i\in[n]$.
For each $i$, the condition $P(\alpha_i)=0$ is equivalent to $\min(\alpha_i,\mathbb F_q)\mid P$. Therefore, $ P(\alpha_i)=0 $, for all $ i \in [n] $, if and only if $ L\mid P $.

If $k\leq \deg(L)$, no nonzero polynomial of
degree less than $k$ can be divisible by $L$. Thus the $k$
generators above are linearly independent and the code has dimension $k$.

If $k>\deg(L)$, evaluation at $\bm\alpha$ depends only on the residue class
of a polynomial modulo $L$. Moreover, every residue class
modulo $L$ has a unique representative of degree less than
$\deg(L)$, and all such representatives already occur among the polynomials
of degree less than $k$. Thus the code has dimension $\deg(L)$.
\end{proof}

\subsection{Linear Hamming and folded isometries} \label{subsec Hamming and folded isometries}

aRS codes are $ \mathbb{F}_q $-linear codes in $ \mathbb{F}_{q^r}^n $. The folded Hamming metric in $ \mathbb{F}_q^{nr} $ and the classical Hamming metric in $ \mathbb{F}_{q^r}^n $ are equivalent, see \cite[Sec. 2]{martinez2025linear}. We revisit some of these results here, and we also characterize the linear isometries between both metrics.

The Hamming weight of $ \mathbf{c} = (c_1,\ldots, c_n) \in \mathbb{F}_{q^r}^n $ is defined as $ \w_H(\mathbf{c}) = | \{ i \in [n] : c_i \neq 0 \} | $. For $ \mathbf{c},\mathbf{d} \in \mathbb{F}_{q^r}^n $, the Hamming distance between them is defined as $ \dd_H(\mathbf{c},\mathbf{d}) = \w_H(\mathbf{c}- \mathbf{d}) $. 

\begin{definition}
Let $\boldsymbol\beta=(\beta_1,\ldots,\beta_r)$ be an ordered $\mathbb F_q$-basis of $\mathbb F_{q^r}$. We define the
expansion map 
$$\varepsilon_{\boldsymbol\beta}:
\mathbb F_{q^r}\longrightarrow\mathbb F_q^{1\times r} \qquad \textrm{by} \qquad 
\varepsilon_{\boldsymbol\beta}
\left(
a_1\beta_1+\cdots+a_r\beta_r
\right)
=
(a_1,\ldots,a_r),
$$
for $ a_1, \ldots, a_r \in \mathbb{F}_q $. We use the same notation for its componentwise extension
\[
\varepsilon_{\boldsymbol\beta}:
\mathbb F_{q^r}^n\longrightarrow\mathbb F_q^{nr}, \qquad \textrm{where} \qquad
\varepsilon_{\boldsymbol\beta}(c_1,\ldots,c_n)
=
\big(
\varepsilon_{\boldsymbol\beta}(c_1)
, \ldots, 
\varepsilon_{\boldsymbol\beta}(c_n)
\big),
\]
for $ c_1, \ldots, c_n \in \mathbb{F}_{q^r} $. More generally, if $\overline{\boldsymbol\beta}
=
(\boldsymbol\beta_1,\ldots,\boldsymbol\beta_n)$ is a collection of ordered $\mathbb F_q$-bases of $\mathbb F_{q^r}$,
we define the blockwise expansion map
\[
\varepsilon_{\overline{\boldsymbol\beta}}:
\mathbb F_{q^r}^n\longrightarrow\mathbb F_q^{nr}, \qquad \textrm{where} \qquad
\varepsilon_{\overline{\boldsymbol\beta}}
(c_1,\ldots,c_n)
=
\big(
\varepsilon_{\boldsymbol\beta_1}(c_1)
, \ldots, 
\varepsilon_{\boldsymbol\beta_n}(c_n)
\big),
\]
for $ c_1, \ldots, c_n \in \mathbb{F}_{q^r} $.
\end{definition}

As remarked in \cite{martinez2025linear}, it is obvious that expansion maps $ \varepsilon_{\overline{\boldsymbol\beta}}:
(\mathbb F_{q^r}^n, \dd_H)\longrightarrow\mathbb (F_q^{nr},\dd_F) $ are $ \mathbb{F}_q $-linear isometries (we indicate the metric corresponding to each space). We now show they are in fact all the linear isometries, up to permutation, between such metrics.

\begin{theorem}\label{th:isometries-extension}
Given an $\mathbb F_q$-linear isometry $\psi:(\mathbb F_{q^r}^n,\dd_H)
\longrightarrow
(\mathbb F_q^{nr},\dd_F)$, there exist a permutation $\sigma\in S_n$ and ordered $\mathbb F_q$-bases
$\boldsymbol\beta_1,\ldots,\boldsymbol\beta_n$ of $\mathbb F_{q^r}$ such that
\[
\psi(c_1,\ldots,c_n)
=
\big(
\varepsilon_{\boldsymbol\beta_1}(c_{\sigma(1)})
, \ldots, 
\varepsilon_{\boldsymbol\beta_n}(c_{\sigma(n)})
\big),
\]
for all $ (c_1, \ldots, c_n) \in \mathbb{F}_{q^r}^n $. Conversely, every map of this form is an $\mathbb F_q$-linear
isometry.
\end{theorem}
\begin{proof}
The converse is obvious. Let now $\psi:(\mathbb F_{q^r}^n,d_H)\longrightarrow
(\mathbb F_q^{nr},d_F)$  be an $\mathbb F_q$-linear isometry, and fix an ordered
$\mathbb F_q$-basis $\boldsymbol\beta$ of $\mathbb F_{q^r}$.
Then $\phi=\psi\circ\varepsilon_{\boldsymbol\beta}^{-1}
:\mathbb F_q^{nr}\longrightarrow\mathbb F_q^{nr}$  is an $\mathbb F_q$-linear folded-Hamming isometry. By Proposition
\ref{prop isometries}, there exist $\sigma\in S_n$ and
$A_1,\ldots,A_n\in\mathrm{GL}_r(\mathbb F_q)$ such that
\[
\phi(\mathbf c_1,\ldots,\mathbf c_n)
=
\big(
\mathbf c_{\sigma(1)}A_1,\ldots,
\mathbf c_{\sigma(n)}A_n
\big),
\]
for $ \mathbf{c}_1, \ldots, \mathbf{c}_n \in \mathbb{F}_q^r $. For $ i \in [n] $, define the $ \mathbb{F}_q $-basis $ \boldsymbol\beta_i = \boldsymbol\beta (A_i^{-1})^\intercal $ of $ \mathbb{F}_{q^r} $. Given $ a \in \mathbb{F}_{q^r} $, then $ \mathbf{a}_i = \varepsilon_{\boldsymbol\beta_i}(a) $ is the unique vector in $ \mathbb{F}_q^r $ such that $ a = \boldsymbol\beta_i \mathbf{a}_i^\intercal $. If $ \mathbf{b} = \varepsilon_{\boldsymbol\beta}(a) $, then
$$ a = \boldsymbol\beta \mathbf{b}^\intercal = \boldsymbol\beta_i A_i^\intercal \mathbf{b}^\intercal. $$
Hence, since $ \mathbf{a}_i = \varepsilon_{\boldsymbol\beta_i}(a) $ and $ \mathbf{b} = \varepsilon_{\boldsymbol\beta}(a) $, uniqueness of the coordinate vector implies that
$$ \varepsilon_{\boldsymbol\beta_i}(a) = \varepsilon_{\boldsymbol\beta}(a) A_i, $$
for all $ a \in \mathbb{F}_{q^r} $ and all $ i \in [n] $. Finally, given $ \mathbf{c}_1, \ldots, \mathbf{c}_n \in \mathbb{F}_q^r $, if we define $ c_i = \varepsilon_{\boldsymbol\beta}^{-1}(\mathbf{c}_i) \in \mathbb{F}_{q^r} $, then
$$ \mathbf{c}_{\sigma(i)} A_i = \varepsilon_{\boldsymbol\beta}(c_{\sigma(i)}) A_i = \varepsilon_{\boldsymbol\beta_i}(c_{\sigma(i)}), $$
for $ i \in [n] $. Therefore,
\begin{equation*}
\begin{split}
\psi(c_1,\ldots, c_n) & = \phi(\mathbf c_1,\ldots,\mathbf c_n)
 \\
  & =
\big(
\mathbf c_{\sigma(1)}A_1,\ldots,
\mathbf c_{\sigma(n)}A_n
\big) = \big(
\varepsilon_{\boldsymbol\beta_1}(c_{\sigma(1)})
, \ldots, 
\varepsilon_{\boldsymbol\beta_n}(c_{\sigma(n)})
\big),
\end{split}
\end{equation*}
and we are done.
\end{proof}

As a corollary of Proposition \ref{prop isometries} and Theorem \ref{th:isometries-extension}, we obtain the following.
 
\begin{corollary} \label{cor isometries hamming}
An $ \mathbb{F}_q $-linear map $ \varphi : \mathbb{F}_{q^r}^n \longrightarrow \mathbb{F}_{q^r}^n $ is an isometry for the Hamming metric if, and only if, there exist a permutation $ \sigma \in S_n $ and $ \mathbb{F}_q $-linear vector space automorphisms $ T_i : \mathbb{F}_{q^r} \longrightarrow \mathbb{F}_{q^r} $, for $ i \in [n] $, such that
$$ \varphi(c_1,\ldots,c_n)
=
\big(
T_1(c_{\sigma(1)})
, \ldots, 
T_n(c_{\sigma(n)})
\big), $$
for $ (c_1, \ldots, c_n) \in \mathbb{F}_{q^r}^n $.
\end{corollary}

As remarked in \cite{martinez2025linear}, we may therefore identify the metric vector spaces $ (\mathbb{F}_{q^r}^n,\dd_H) $ and $ (\mathbb{F}_q^{nr}, \dd_F) $, and all of the concepts and results valid for the second can be directly translated to the first one. In particular, the definitions of QMDS and dually QMDS codes can be directly translated to $ (\mathbb{F}_{q^r}^n,\dd_H) $. 

We may also define additive generalized Reed--Solomon (aGRS) codes by considering all codes equivalent to aRS codes. This extends \cite[Sec. 4]{yadav}.

\begin{definition}
Let $\bm\alpha=(\alpha_1,\ldots,\alpha_n)\in\mathbb F_{q^r}^n$ and, for $i\in[n]$, let $T_i:\mathbb F_{q^r}\longrightarrow\mathbb F_{q^r}$ be $\mathbb F_q$-linear vector space automorphisms. We define the additive generalized Reed--Solomon (aGRS) code 
\[
\mathrm{aGRS}_q(\bm\alpha,k;T_1,\ldots,T_n)
=
\Big\{
\big(T_1(P(\alpha_1)),\ldots,T_n(P(\alpha_n))\big):
P\in\mathbb F_q[X],\ \deg(P)<k
\Big\}.
\]
\end{definition}

\subsection{The intersection between the aRS and PI families} \label{subsec intersection}

We now show that aRS codes with nonconjugate evaluation points are equivalent or isometric (see Theorem \ref{th:isometries-extension}) to those PI codes defined by irreducible polynomials. Consequently, as pointed out at the end of Subsection \ref{subsec Hamming and folded isometries}, every property invariant under
$\mathbb F_q$-linear isometries is shared by both families of codes.

\begin{theorem}\label{PIC}
Let $P_1,\ldots,P_n\in\mathbb F_q[X]$ be distinct monic irreducible polynomials
of degree $r$.
For each $i\in[n]$, choose a root
$\alpha_i\in\mathbb F_{q^r}$ of $P_i$ and let $\boldsymbol\beta_i=(1,\alpha_i,\ldots,\alpha_i^{r-1})$ and $\overline{\boldsymbol\beta}
=
(\boldsymbol\beta_1,\ldots,\boldsymbol\beta_n)$.
Then each $\boldsymbol\beta_i$ is an ordered $\mathbb F_q$-basis of
$\mathbb F_{q^r}$ and for
$\bm\alpha=(\alpha_1,\ldots,\alpha_n)$, we have that
\[
\varepsilon_{\overline{\boldsymbol\beta}}
\big(\mathrm{aRS}_q(\bm\alpha,k)\big)
=
\mathrm{PI}_k(P_1,\ldots,P_n).
\]

Conversely, if $\alpha_1,\ldots,\alpha_n\in\mathbb F_{q^r}$ belong to
pairwise distinct conjugacy classes of size $r$, then the same holds with $P_i=\min(\alpha_i,\mathbb F_q)$, $\boldsymbol\beta_i=(1,\alpha_i,\ldots,\alpha_i^{r-1})$, for $ i \in [n] $, 
and $
\overline{\boldsymbol\beta}
=
(\boldsymbol\beta_1,\ldots,\boldsymbol\beta_n)$.
\end{theorem}
\begin{proof}
Since $P_i$ is irreducible in $ \mathbb{F}_q[X] $ of degree $r$ and $\alpha_i$ is a root of
$P_i$, we have $P_i=\min(\alpha_i,\mathbb F_q)$. Hence $\boldsymbol\beta_i=(1,\alpha_i,\ldots,\alpha_i^{r-1})$ 
is an ordered $\mathbb F_q$-basis of
$\mathbb F_q(\alpha_i)=\mathbb F_{q^r}$, for $ i \in [n] $. Moreover, since the polynomials $ P_i = \min(\alpha_i,\mathbb{F}_q) $ are distinct of degree $ r $, the roots $ \alpha_i $ belong to pairwise distinct conjugacy classes of size $ r $.

Now, given $P\in\mathbb F_q[X]$ with $\deg(P)<k$, write $\varphi_i(P)=(c_{i,0},\ldots,c_{i,r-1}) \in \mathbb{F}_q^r $. By definition,
\[
P(X)\equiv
c_{i,0}+c_{i,1}X+\cdots+c_{i,r-1}X^{r-1}
\mod{P_i(X)}.
\]
Since $P_i(\alpha_i)=0$, evaluating at $\alpha_i$ gives $P(\alpha_i)
=
c_{i,0}+c_{i,1}\alpha_i+\cdots+c_{i,r-1}\alpha_i^{r-1}$ and therefore $\varepsilon_{\boldsymbol\beta_i}\big(P(\alpha_i)\big)
=
\varphi_i(P)$ for all $i\in[n]$. Applying this blockwise yields
\[
\varepsilon_{\overline{\boldsymbol\beta}}
\big(P(\alpha_1),\ldots,P(\alpha_n)\big)
=
\big(\varphi_1(P), \ldots, \varphi_n(P)\big).
\]
Letting $P$ range over all polynomials of degree less than $k$, we obtain
\[
\varepsilon_{\overline{\boldsymbol\beta}}
\big(\mathrm{aRS}_q(\bm\alpha,k)\big)
=
\mathrm{PI}_k(P_1,\ldots,P_n).
\]

For the converse, if $\alpha_1,\ldots,\alpha_n$ belong to pairwise
distinct conjugacy classes of size $r$, then $P_i=\min(\alpha_i,\mathbb F_q)$ for $i\in[n]$
are distinct irreducible polynomials of degree $r$. Hence the preceding
argument applies with $\boldsymbol\beta_i=(1,\alpha_i,\ldots,\alpha_i^{r-1})$, 
which proves the converse.
\end{proof}

In particular, from the definitions and Proposition \ref{prop isometries}, Theorem \ref{th:isometries-extension} and Corollary \ref{cor isometries hamming}, we obtain the same result regarding generalized codes.

\begin{corollary}\label{GPI aGRS}
Every code $\mathrm{GPI}_k(P_1,\ldots,P_n;A_1,\ldots,A_n) \subseteq \mathbb{F}_q^{nr} $, where $P_1,\ldots,P_n\in\mathbb F_q[X]$ are monic irreducible polynomials
of degree $r$, is equivalent to a code $\mathrm{aGRS}_q(\bm\alpha,k;T_1,\ldots,T_n) \subseteq \mathbb{F}_{q^r}^n $ such that $\alpha_1,\ldots,\alpha_n\in \mathbb{F}_{q^r}$ belong to pairwise distinct conjugacy classes of size $r$ and conversely, where $P_i=\min(\alpha_i,\mathbb F_q)$ for all $i\in[n]$.
\end{corollary}

Now from Theorems \ref{thm:PIC-dual} and \ref{PIC} and Corollary \ref{GPI aGRS}, we obtain that the dual of an aRS (and in general an aGRS) code is again an aGRS code, translating the inner product in $ \mathbb{F}_q^{nr} $ to $ \mathbb{F}_{q^r}^n $. We will show that this translated product is the trace inner product (using the right bases). Denote by $ {\rm Tr}_q $ the trace of $ \mathbb{F}_{q^r} $ over $ \mathbb{F}_q $.

\begin{definition} \label{def trace dual}
Given an $ \mathbb{F}_q $-linear code $ \mathcal{C} \subseteq \mathbb{F}_{q^r}^n $, and denoting by $ \cdot $ the standard inner product in $ \mathbb{F}_{q^r}^n $, we define its dual with respect to the trace inner product
$$\mathcal{C}^\perp = \{ \mathbf{d} \in \mathbb{F}_{q^r}^n : {\rm Tr}_q(\mathbf{c} \cdot \mathbf{d}) = 0, \textrm{ for all } \mathbf{c} \in \mathcal{C} \} .$$
\end{definition}

The proof of the following proposition is straightforward.

\begin{proposition} \label{prop inner products}
Let $ (\beta_1, \ldots, \beta_r) , (\gamma_1, \ldots, \gamma_r) \in \mathbb{F}_{q^r}^r $ be ordered basis that are dual of each other (i.e., $ {\rm Tr}(\beta_i \gamma_j) = \delta_{i,j} $), which always exist, see \cite[p. 54]{lidl}. Then for $ \mathbf{c},\mathbf{d} \in \mathbb{F}_{q^r}^n $, we have
$$ {\rm Tr}_q( \mathbf{c} \cdot \mathbf{d}) = \varepsilon_{\boldsymbol\beta}(\mathbf{c}) \cdot  \varepsilon_{\boldsymbol\gamma}(\mathbf{d}) . $$
In particular, for an $ \mathbb{F}_q $-linear code $ \mathcal{C} \subseteq \mathbb{F}_{q^r}^n $, we have
$$ \varepsilon_{\boldsymbol\beta}( \mathcal{C}^\perp) = \varepsilon_{\boldsymbol\gamma}(\mathcal{C})^\perp. $$
\end{proposition}

Thus we deduce from Theorems \ref{thm:PIC-dual}, \ref{PIC} and \ref{GPI aGRS} that the dual of an aRS code is an aGRS code, which was already obtained in \cite[Th. 4.3]{yadav}. However, we give a more explicit formula (by giving the multipliers) and a shorter proof. The explicit version will be useful when we compute the permutation group of aRS codes. We will need the following ``subfield Lagrange interpolation'' result. 

\begin{lemma}  \label{lemma lagrange}
Let $\alpha_1,\ldots,\alpha_n \in \mathbb{F}_{q^r}$ belong to pairwise distinct
conjugacy classes of size $r$, and let $ L = \prod_{i=1}^n \min(\alpha_i,\mathbb{F}_q) \in \mathbb{F}_q[X] $. For all $ P \in \mathbb{F}_q[X] $ with $ \deg(P) < nr $,  
$$ P(X) = \sum_{i=1}^n {\rm Tr}_q \left( P(\alpha_i) \frac{L(X)}{(X-\alpha_i) L^\prime(\alpha_i)} \right) , $$
where $ {\rm Tr}_q(a_0 + \cdots + a_k X^k) = {\rm Tr}_q(a_0) + \cdots + {\rm Tr}_q(a_k) X^k $, for $ a_0, \ldots, a_k \in \mathbb{F}_{q^r} $.
\end{lemma}
\begin{proof}
Consider the classical Lagrange interpolation formula in $ \mathbb{F}_{q^r}[X] $, for the $ nr $ evaluation points $ \alpha_i^{q^j} $, for $ j \in [0,r-1] $ (which give the full conjugacy class of $ \alpha_i $), for $ i \in [n] $. Since $ L(X) = \prod_{i=1}^n\prod_{j=0}^{r-1} (X-\alpha_i^{q^j}) \in \mathbb{F}_q[X] $, the formula is
$$ P(X) = \sum_{i=1}^n \sum_{j=0}^{r-1} P(\alpha_i)^{q^j} \frac{L(X)}{\left(X-\alpha_i^{q^j}\right) L^\prime(\alpha_i^{q^j})} , $$
which is the same formula as in the statement.
\end{proof}

\begin{proposition} \label{prop:aRS-dual-full-degree}
Let $\mathcal C=\mathrm{aRS}_q(\bm\alpha,k) \subseteq \mathbb{F}_{q^r}^n $, where
$\alpha_1,\ldots,\alpha_n \in \mathbb{F}_{q^r}$ belong to pairwise distinct
conjugacy classes of size $r$. Let $ L = \prod_{i=1}^n \min(\alpha_i,\mathbb{F}_q) \in \mathbb{F}_q[X] $ and $ \lambda_i = L^\prime (\alpha_i)^{-1} \in \mathbb{F}_{q^r}^* $, for $ i \in [n] $. Then
$$ \mathcal{C}^\perp = \{ (\lambda_1 Q(\alpha_1) , \ldots, \lambda_n Q(\alpha_n) ) : Q \in \mathbb{F}_q[X], \deg(Q) < nr-k \}, $$
which is the code $ {\rm aGRS}_q(\boldsymbol\alpha,nr-k;T_1,\ldots,T_n) $, where $ T_i : \mathbb{F}_{q^r} \longrightarrow \mathbb{F}_{q^r} $ is the $ \mathbb{F}_q $-linear automorphism given by $ T_i(c) = \lambda_i c $, for all $ c \in \mathbb{F}_{q^r} $ and all $ i \in [n] $.
\end{proposition}
\begin{proof}
Let $ P,Q \in \mathbb{F}_q[X] $ with $ \deg(P) < k $ and $ \deg(Q) < nr- k $. Then $ \deg(PQ) < nr-1 $. By Lemma \ref{lemma lagrange}, 
$$ P(X)Q(X) = \sum_{i=1}^n {\rm Tr}_q \left( P(\alpha_i)Q(\alpha_i) \frac{L(X)}{(X-\alpha_i) L^\prime(\alpha_i)} \right) . $$
Now, since $ \deg(PQ) < nr-1 $, $ \operatorname{coef}(PQ,nr-1) = 0 $. Since $ L(X) $ is monic, we deduce that
$$ 0 = \operatorname{coef}(PQ,nr-1) = \sum_{i=1}^n {\rm Tr}_q \left( \frac{P(\alpha_i)Q(\alpha_i)}{L^\prime(\alpha_i)} \right) = \sum_{i=1}^n {\rm Tr}_q ( \lambda_i P(\alpha_i)Q(\alpha_i)) . $$
Thus 
$$ \{ (\lambda_1 Q(\alpha_1) , \ldots, \lambda_n Q(\alpha_n) ) : Q \in \mathbb{F}_q[X], \deg(Q) < nr-k \} \subseteq \mathcal{C}^\perp , $$
and since both have the same dimension, they must be equal.
\end{proof}

We deduce that aRS codes (and thus aGRS codes) are dually QMDS, which was proven in \cite[Th. 4.3]{yadav}.

\begin{corollary} [\cite{yadav}] \label{cor aRS are dually}
If $ k \in [0,nr] $ and $\alpha_1,\ldots,\alpha_n \in \mathbb{F}_{q^r}$ belong to pairwise distinct
conjugacy classes of size $r$, then $ \mathrm{aRS}_q(\bm\alpha,k) \subseteq \mathbb{F}_{q^r}^n $ is dually QMDS.
\end{corollary}

As a consequence of Corollary \ref{cor aRS are dually} and \cite[Th. 4]{martinez2025linear}, we obtain the weight distribution of aRS codes (and thus also aGRS codes). By the discussion in Subsection \ref{subsec Hamming and folded isometries}, we may consider the classical Hamming weight distribution.

\begin{theorem}\label{thm:aRS-weight-distribution}
Let $\mathcal C=\mathrm{aRS}_q(\bm\alpha,k) \subseteq \mathbb{F}_{q^r}^n $ with $\alpha_1,\ldots,\alpha_n \in \mathbb{F}_{q^r} $ in pairwise distinct conjugacy classes of size $r$, and let $(W_j)_{j=0}^n$ be the Hamming weight distribution of $ \mathcal{C} $. Then $W_0=1$, $W_j=0$ for $j \in [d-1]$, and for $j \in [d, n]$,
\[
W_j
=
\binom nj
\sum_{i=0}^{j-d}
(-1)^i\binom ji
\big(q^{k-r(n-j+i)}-1\big),
\qquad \textrm{where }
d=n-\left\lceil\frac{k}{r}\right\rceil+1.
\]
\end{theorem}

Finally, among aRS codes whose evaluation points lie in pairwise distinct conjugacy classes of size $r$, the largest possible length is exactly the number of such conjugacy classes. 

\begin{corollary}\label{cor:max-full-degree-aRS}
The maximum possible length of an aRS code $\mathrm{aRS}_q(\bm\alpha,k) \subseteq \mathbb{F}_{q^r}^n$, where $\alpha_1,\ldots,\alpha_n $ $ \in \mathbb{F}_{q^r} $ belong to pairwise distinct conjugacy classes of size $r$, is $N_q(r)$. It is obtained by choosing representatives $ \alpha_i $ of all of the conjugacy classes of size $ r $.
\end{corollary}

Combining this with Corollaries \ref{cor:asymptotically} and \ref{cor:asymptotic-Mq}, we obtain, for fixed $q \geq 2$ and $ r \geq 4 $, respectively,
\[
\lim_{r\to\infty}\frac{N_q(r)}{M_q(r)}=1 \quad \textrm{and} \quad \lim_{q\to\infty}\frac{N_q(r)}{M_q(r)}=1.
\]
Thus, aRS codes whose evaluation points lie in pairwise distinct
conjugacy classes of size $r$ asymptotically attain the maximum
possible length of PI codes.

\subsection{The permutation group of aRS codes} \label{subsec permutation}
We now study the coordinate permutations preserving an aRS code. By Corollary \ref{cor isometries hamming}, they form a subgroup of $ \mathbb{F}_q $-linear isometries for the Hamming metric that leave the aRS code invariant.

\begin{definition}
Let $\mathcal C\subseteq\mathbb F_{q^r}^n$ be a code. For all
$\sigma\in S_n$, define $\Pi_\sigma:\mathbb F_{q^r}^n\longrightarrow\mathbb F_{q^r}^n$ by $\Pi_\sigma(c_1,\ldots,c_n)
=
(c_{\sigma(1)},\ldots,c_{\sigma(n)})$, for $ (c_1,\ldots, c_n) \in \mathbb{F}_{q^r}^n $.
The permutation group of $\mathcal C$ is
\[
\operatorname{PAut}(\mathcal C)
=
\{\sigma\in S_n:\Pi_\sigma(\mathcal C)=\mathcal C\}.
\]
\end{definition}

Setting $ r=1 $, the result below recovers the corresponding characterization for classical Reed--Solomon codes over arbitrary evaluation points recently obtained in \cite{camps}. The permutation group of classical Reed--Solomon codes evaluated over the whole field or its multiplicative group was originally obtained in \cite{dur1987}.

\begin{theorem}\label{thm:permutation-group-aRS}
Let $\mathcal C=\mathrm{aRS}_q(\bm\alpha,k)
\subseteq\mathbb F_{q^r}^n$, with $2\leq k\leq nr-2$, and assume that
$\alpha_1,\ldots,\alpha_n$ $ \in \mathbb{F}_{q^r} $ belong to pairwise distinct
conjugacy classes of size $r$. For a permutation $ \sigma \in S_n $, we have that $ \sigma \in \operatorname{PAut}(\mathcal C) $ if and only if there exist $ a\in\mathbb F_q^* $ and $ b\in\mathbb F_q $ such that $ \alpha_{\sigma(i)}=a\alpha_i+b $, for all $ i\in[n] $.
\end{theorem}

\begin{proof}
First, suppose that there exist
$a\in\mathbb F_q^*$ and $b\in\mathbb F_q$ such that
$\alpha_{\sigma(i)}=a\alpha_i+b$ for all $i\in[n]$. For all
$P\in\mathbb F_q[X]$ with $\deg(P)<k$,
\[
\Pi_\sigma\big(P(\alpha_1),\ldots,P(\alpha_n)\big)
=
\big(P(a\alpha_1+b),\ldots,P(a\alpha_n+b)\big).
\]
Since $P(aX+b)\in\mathbb F_q[X]$ and
$\deg(P(aX+b))<k$, it follows that
$\Pi_\sigma(\mathcal C)\subseteq\mathcal C$. As $\Pi_\sigma$ is an
isomorphism, we obtain $\Pi_\sigma(\mathcal C)=\mathcal C$, and hence
$\sigma\in\operatorname{PAut}(\mathcal C)$.

We now prove the converse. For all $i\in[n]$, let
$P_i(X)=\min(\alpha_i,\mathbb F_q)$ and set
$L(X)=\prod_{i=1}^nP_i(X)$. By assumption, the polynomials
$P_1,\ldots,P_n$ are distinct irreducible polynomials in $ \mathbb{F}_q[X] $ of degree $r$,
and hence $\deg(L)=nr$.

Let $\sigma\in\operatorname{PAut}(\mathcal C)$. Since $k\geq2$, we have that
$(\alpha_1,\ldots,\alpha_n)\in\mathcal C$ and therefore there exists
$h_\sigma\in\mathbb F_q[X]$ with $\deg(h_\sigma)<k$ such that
$h_\sigma(\alpha_i)=\alpha_{\sigma(i)}$ for all $i\in[n]$.
Since the $\alpha_i$ are pairwise distinct and have degree $ r $ over $ \mathbb{F}_q $, it must be
\begin{equation}
 \deg(h_\sigma) \geq 1.
\label{eq deg(h) geq 1}
\end{equation}

Moreover, for all $j\in[k-1]$, we see that $\big(h_\sigma^j(\alpha_1),\ldots,h_\sigma^j(\alpha_n)\big)
=
\Pi_\sigma\big(\alpha_1^j,\ldots,\alpha_n^j\big)
\in\mathcal C$. 
Hence there exists $f_j\in\mathbb F_q[X]$ with $\deg(f_j)<k$ such that
$h_\sigma^j(\alpha_i)=f_j(\alpha_i)$ for all $i\in[n]$. Thus
\begin{equation}
 h_\sigma^j\equiv f_j\mod L.
\label{eq h sigma j cong fj}
\end{equation}

Assume first that $2k-2<nr$. We prove inductively in $ j \in [k-1] $ that
$h_\sigma^j=f_j$. For $j=1$, by definition we have that $f_1=h_\sigma$.
If $h_\sigma^{j-1}=f_{j-1}$ for some $ j \in [2, k-1] $, then
\[
\deg(h_\sigma^j)
=\deg(h_{\sigma}^{j-1})+\deg(h_\sigma)=
\deg(f_{j-1})+\deg(h_\sigma)
\leq2k-2<nr=\deg(L).
\]
Since also $\deg(f_j)<k<nr$, the congruence
$h_\sigma^j\equiv f_j\mod L$ implies $h_\sigma^j=f_j$.
In particular,
$(k-1)\deg(h_\sigma)=\deg(h_\sigma^{k-1})<k$, and therefore
$\deg(h_\sigma)\leq1$.

Suppose now that $nr\leq2k-2$, and set $k^\prime=nr-k$. By the assumption
$k\leq nr-2$, we have $k^\prime\geq2$. By Proposition \ref{prop:aRS-dual-full-degree}, 
$$ \mathcal{C}^\perp = \{ (\lambda_1 Q(\alpha_1) , \ldots, \lambda_n Q(\alpha_n) ) : Q \in \mathbb{F}_q[X], \deg(Q) < k^\prime \}, $$
where $ \lambda_i = L^\prime(\alpha_i)^{-1} \in \mathbb{F}_{q^r}^* $. It is straightforward to verify that $ \sigma \in \operatorname{PAut}(\mathcal C^\perp) $. 

Applying $ \Pi_\sigma $ to the codewords of $ \mathcal{C}^\perp $ corresponding to $Q=1$ and $Q=X$, we obtain $g_0,g_1\in\mathbb F_q[X]$ with
$\deg(g_0),\deg(g_1)<k^\prime$ such that $\lambda_{\sigma(i)}
=
\lambda_i g_0(\alpha_i)$
and
$\lambda_{\sigma(i)}\alpha_{\sigma(i)}
=
\lambda_i g_1(\alpha_i)
$
for all $i\in[n]$. Since
$\alpha_{\sigma(i)}=h_\sigma(\alpha_i)$, we obtain
$g_0(\alpha_i)h_\sigma(\alpha_i)=g_1(\alpha_i)$ for all $i \in [n]$.
Therefore
$g_0h_\sigma\equiv g_1\mod L$.
Now
$\deg(g_0h_\sigma)\leq(k^\prime -1)+(k-1)=nr-2$ and
$\deg(g_1)<k^\prime <nr$, so $g_0h_\sigma=g_1$.
Since $g_0\neq0$, it follows that $\deg(h_\sigma)<k^\prime $.

We now return to the congruences $h_\sigma^j\equiv f_j\pmod L$ from (\ref{eq h sigma j cong fj}), for $ j \in [k-1] $. Again, taking $f_1=h_\sigma$, if
$h_\sigma^{j-1}=f_{j-1}$, then
\[
\deg(h_\sigma^j)
=
\deg(f_{j-1})+\deg(h_\sigma)
<
k+k^\prime
=
nr.
\]
Since $\deg(f_j)<k<nr$, we obtain $h_\sigma^j=f_j$. Hence
$(k-1)\deg(h_\sigma)<k$, and again $\deg(h_\sigma)\leq1$.

Thus, in both cases $ 2k-2 < nr $ and $ nr \leq 2k-2 $, we have that $ \deg(h_\sigma) \leq 1 $. Since $ \deg(h_\sigma) \geq 1 $ by (\ref{eq deg(h) geq 1}), we deduce that $ \deg(h_\sigma) = 1 $. In other words, $h_\sigma(X)=aX+b$ for some
$a\in\mathbb F_q^*$ and $b\in\mathbb F_q$, and
$\alpha_{\sigma(i)}=a\alpha_i+b$ for all $i\in[n]$.
\end{proof}

In particular, we obtain the following.

\begin{corollary}
Let $ \mathcal{A} = \{ \alpha_1, \ldots, \alpha_n \} \subseteq \mathbb{F}_{q^r} $, where $ \alpha_1, \ldots, \alpha_n $ belong to pairwise distinct conjugacy classes of size $ r $. Define $ \operatorname{AGL}_q(\mathcal{A}) $ as the group of maps $ f : \mathcal{A} \longrightarrow \mathcal{A} $ such that there exist $ a \in \mathbb{F}_q^* $ and $ b \in \mathbb{F}_q $ satisfying $ f(\alpha) = a \alpha +b $, for all $ \alpha \in \mathcal{A} $. If $ 2 \leq k \leq nr-2 $, there is a group isomorphism
$$ \operatorname{PAut}(\mathrm{aRS}_q(\bm\alpha,k))
\cong \operatorname{AGL}_q(\mathcal{A}). $$
\end{corollary}

\subsection{Additive Reed--Solomon codes with arbitrary evaluation points}\label{subsec aRS not covered}

In \cite[Th. 4.1]{yadav} the authors characterize when an aRS code with different evaluation points is QMDS. In this subsection, we characterize when aRS codes over arbitrary evaluation points $ \alpha_1, \ldots, \alpha_n \in \mathbb{F}_{q^r} $ (different or not) are QMDS and/or dually QMDS. In particular, we will show that the dually QMDS aRS codes from Subsection \ref{subsec intersection} attain the maximum length among all QMDS aRS codes, except possibly by one unit. The ideas are similar to those in Subsection \ref{subsec PR codes}, but we write full proofs for convenience of the reader that is only interested in aRS codes.

Throughout this subsection, fix $ k \in [0,nr] $ and $ \mathcal{C} =\operatorname{aRS}_q(\boldsymbol\alpha,k)\subseteq\mathbb F_{q^r}^n$,
where $\boldsymbol\alpha=(\alpha_1,\ldots,\alpha_n) $ $ \in \mathbb{F}_{q^r}^n $ is arbitrary, and define
\[
 P_i=\min(\alpha_i,\mathbb F_q),\qquad d_i=\deg (P_i),\qquad \textrm{and} \qquad
 L_J=\operatorname{lcm}\{P_j : j\in J \},
\]
for $ i \in [n] $ and $ J \subseteq [n] $, where we define $L_\varnothing=1$ and $ L = L_{[n]} $. By Proposition \ref{prop:dim-aRS}, replacing the degree
parameter by $\min\{k,\deg (L) \}$ leaves the code unchanged.
Thus, throughout this subsection, we may and do assume
$k=\dim(\mathcal{C}) \leq \deg(L) $.
 
We start by looking at punctured and shortened codes. The next lemma has essentially the same proof as \cite[Th. 1.5.7]{pless}. Recall that we are considering trace duals in $ \mathbb{F}_{q^r}^n $ (Definition \ref{def trace dual}).

\begin{lemma} \label{lemma punctured shortened}
For $ J \subseteq [n] $, an $ \mathbb{F}_q $-linear code $ \mathcal{D} \subseteq \mathbb{F}_{q^r}^n $, and $ \pi_J : \mathbb{F}_{q^r}^n \longrightarrow \mathbb{F}_{q^r}^{|J|} $ the projection onto the positions indexed by $ J $, define the punctured and shortened codes of $ \mathcal{D} $ at $ J $, respectively, as
$$ \mathcal{D}^J = \pi_J(\mathcal{D}) \qquad \textrm{and} \qquad \mathcal{D}_J = \pi_J(\mathcal{D} \cap \ker(\pi_{[n] \setminus J})). $$
Then we have that $ (\mathcal{D}^J)^\perp = (\mathcal{D}^\perp)_J $ and $ (\mathcal{D}_J)^\perp = (\mathcal{D}^\perp)^J $.
\end{lemma}

\begin{lemma}\label{lem:ars-projections}
For $J\subseteq[n]$, $ \dim(\mathcal{C}^J)=\min\{k,\deg (L_J)\} $ and $ \dim((\mathcal{C}^\perp)_J)= |J|r - \min\{k,\deg (L_J)\} $.
\end{lemma}
\begin{proof}
The code $ \C^J $ is the aRS code with degree restriction $ k $ and over the evaluation set $ \alpha_j $, for $ j \in J $. Thus one can show that $ \dim(\mathcal{C}^J) = \min\{k,\deg (L_J)\} $ exactly as in Proposition \ref{prop:dim-aRS}. Now, since $ (\mathcal{C}^\perp)_J = (\mathcal{C}^J)^\perp $ by Lemma \ref{lemma punctured shortened}, we deduce the formula for $ \dim((\mathcal{C}^\perp)_J) $. 
\end{proof}

The next result characterizes the QMDS property without requiring
the evaluation points to be distinct or to have degree $r$ over $ \mathbb{F}_q $. It extends \cite[Th. 4.1]{yadav} to arbitrary evaluation points (not necessarily distinct) and gives a characterization only in terms of the parameters of the code and the degrees $ d_1,\ldots, d_n $, and not of the subsets $ J \subseteq [n] $. 

\begin{theorem}\label{thm:ars-qmds-arbitrary}
Assume $k>0$, and set $s=\lceil k/r\rceil$. The following hold:
\begin{enumerate}
\item If $k\le r$, then $\mathcal{C}$ is QMDS if and only if $d_i\ge k$
for all $i\in[n]$.
\item \label{item k>r qmds ars}
If $k>r$, then $\mathcal{C}$ is QMDS if and only if the $P_i$ are
pairwise distinct, all but at most one of the $d_i$ equal $r$, and
any exceptional degree satisfies
\[
 d_i\ge k-(s-1)r.
\]
\end{enumerate}
\end{theorem}
\begin{proof}
A nonzero codeword in $\mathcal{C}$ vanishes on the coordinates in $J$ if and
only if $ \dim_{\mathbb{F}_q}(\mathcal{C}^J)<k$. Hence by the Singleton bound (Proposition \ref{singleton}, Item \ref{two}), $ \mathcal{C} $ is QMDS if and only if $ \dim_{\mathbb{F}_q}(\mathcal{C}^J)=k$, for all $ J \subseteq [n] $ such that $ |J| = s $. By Lemma \ref{lem:ars-projections}, this holds if and only if $ \deg(L_J) \geq k $, for all $ J \subseteq [n] $ such that $ |J| = s $.

If $s=1$ (i.e., $ k \leq r $), such a condition is clearly equivalent to $d_i\ge k$, for all $i \in [n]$.
Assume now that $s\ge2$ (i.e., $ k > r $). Assume also that $P_i=P_j$ for $ i,j \in [n] $ with $ i \neq j $. Then for any $ J \subseteq [n] $ with $ i,j \in J $ and $ |J|=s $, we obtain
$$ \deg(L_J) \leq (s-1)r < k , $$
which implies that $ \mathcal{C} $ is not QMDS. Assume then that $ P_1, \ldots, P_n $ are distinct. If there exist $ i,j \in [n] $ with $ i \neq j $ and $ \max\{ d_i, d_j \} < r $, then it must be $ \max \{ d_i,d_j \} \leq r/2 $, since $ d_i $ and $ d_j $ both divide the extension degree $ r $. Thus if $ J \subseteq [n] $ is such that $ i,j \in J $ and $ |J|=s $, then
$$ \deg(L_J) \leq (s-2)r + d_i+d_j \leq (s-1)r < k . $$
Therefore if $ \mathcal{C} $ is QMDS, then there exists at most one index $ i \in [n] $ such that $ d_i < r $. Now if we choose $ J \subseteq [n] $ with $ i \in J $ and $ |J|=s $, then the condition
$$ k \leq \deg(L_J) \leq (s-1)r + d_i $$
implies that $ d_i \geq k - (s-1) r $. Thus if $ \mathcal{C} $ is QMDS (and $ k > r $), the condition in Item \ref{item k>r qmds ars} follows. Assume finally that this condition holds (and $ k > r $). Then given $ J \subseteq [n] $ with $ |J|=s $, we clearly have that $ \deg(L_J) = rs \geq k $ if $ i \notin J $, and 
$$ \deg(L_J) = (s-1)r + d_i \geq k $$
if $ i \in J $, and therefore $ \mathcal{C} $ is QMDS.
\end{proof}

Among QMDS aRS codes (Theorem \ref{thm:ars-qmds-arbitrary} above), those with pairwise nonconjugate evaluation points (i.e., those dually QMDS aRS codes from Subsection \ref{subsec intersection}) attain the maximum possible length except possibly by one unit. This follows from Corollary \ref{cor:max-full-degree-aRS} and the following result, which follows immediately from Theorem \ref{thm:ars-qmds-arbitrary}.

\begin{corollary}\label{cor:ars-qmds-maxlength}
The largest length of an $ \mathbb{F}_q $-linear QMDS aRS code in $ \mathbb{F}_{q^r}^n $ of dimension $ k > r $ is $ N_q(r)+1 $, if $ k \leq N_q(r)+1 $ and $ d \geq k-(s-1)r $ for some proper divisor $ d $ of $ r $, and is $ N_q(r) $ otherwise (again provided that $ k \leq N_q(r) $).
\end{corollary}

Next we look at the dual $ \mathcal{C}^\perp $, also for arbitrary evaluation points.

\begin{theorem}\label{thm:ars-dual-distance}
With the convention $d(\{ \mathbf{0} \})=n+1$, we have
\[
 d(\mathcal{C}^\perp)=
 \begin{cases}
 1,&\text{if some $d_i<r$,}\\
 \min\{2,\lfloor k/r\rfloor+1\},
   &\text{if every $d_i=r$ and some $P_i=P_j$ with $i\ne j$,}\\
 \lfloor k/r\rfloor+1,
   &\text{if $P_1, \ldots, P_n$ are pairwise distinct and all have degree $r$.}
 \end{cases}
\]
In particular, setting $t=\lfloor k/r\rfloor$, $\mathcal{C}^\perp$ is
QMDS if and only if
\begin{enumerate}
\item $t=0$ (i.e., $ 0 \leq k < r $); or
\item $t=1$ (i.e., $ r \leq k < 2r $) and every $d_i=r$; or
\item $t\ge2$ (i.e., $ 2r \leq k $) and the $P_i$ are pairwise distinct of degree $r$.
\end{enumerate}
\end{theorem}
\begin{proof}
By Lemma \ref{lem:ars-projections}, there exists a nonzero codeword in $ \mathcal{C}^\perp $ that vanishes outside of a subset $ J \subseteq [n] $ if and only if $ \min \{k, \deg(L_J) \} < |J|r $.

Assume there exists an $ i \in [n] $ such that $ d_i < r $. Choosing $ J = \{ i \} \subseteq [n] $, we deduce that there is a nonzero codeword in $ \mathcal{C}^\perp $ that vanishes outside of $ J $, thus $ \dd(\mathcal{C}^\perp) = 1 $ in this case.

Assume now that $d_i=r$, for all $ i \in [n] $.
If $k<r$, then $ \deg(L_J) = r $ for all $ J \subseteq [n] $ with $ |J| = 1 $, and therefore $\dd(\mathcal{C}^\perp)=1$. Assume for the rest of the proof that $k\ge r$. Now we deduce similarly that $ \dd(\mathcal{C}^\perp) \geq 2 $. 

If $P_i=P_j$ for some $ i,j \in [n] $ such that
$i\ne j$, then for $ J = \{ i,j \} $ we have that $ \min \{ k,\deg(L_J)\} $ $ = r < 2r $, thus there exists a nonzero codeword in $ \mathcal{C}^\perp $ that vanishes outside of $ J $, giving $d(\mathcal{C}^\perp)=2$.

Finally, if $P_1, \ldots, P_n$ are pairwise distinct, then
$\deg (L_J)=r|J|$ for all $J \subseteq [n]$. Therefore, there exists a nonzero codeword in $ \mathcal{C}^\perp $ that vanishes outside of $ J $ if and only if $ r|J| > k $, i.e., $ |J| \geq \lfloor k/r\rfloor + 1 $. Thus $\dd(\mathcal{C}^\perp)=\lfloor k/r\rfloor+1$,
including the case $k=nr$ by the zero-code convention.
The QMDS criterion follows from the dual Singleton bound (Proposition \ref{singleton}, Item \ref{three}).
\end{proof}

We conclude with a characterization of when arbitrary aRS codes are dually QMDS. Notice that, for $ k > r $, these are exactly the aRS codes from Subsection \ref{subsec intersection}. The proof is straightforward using Theorems \ref{thm:ars-qmds-arbitrary} and \ref{thm:ars-dual-distance}.

\begin{corollary}\label{cor:ars-dually-qmds-arbitrary}
The aRS code $\mathcal{C}$ is dually QMDS if and only if
\begin{enumerate}
\item $0\le k\le r$ and $d_i\ge k$ for all $i\in[n]$; or
\item $r<k\le nr$ and $P_1, \ldots, P_n$ are pairwise distinct of degree $r$.
\end{enumerate}
\end{corollary}

\section*{Acknowledgements}

This work has been supported by MICIU/AEI/ 10.13039/501100011033 and ERDF/EU (Grant no. PID2022-138906NB-C21), and by the Department of Education of the Junta de Castilla y León and FEDER Funds (Reference: CLU-2025-1-02- IMUVA).

\bibliographystyle{plain}

\end{document}